\documentclass[useAMS,usenatbib]{mn2e}

\usepackage{graphicx} 
\usepackage{rotating}
\usepackage{subfig}

\def\HI{H\,{\sc i}}
\def\kms{~km\,s$^{-1}$}

\def\MHI{$M_{\rm HI}$}
\def\Msun{~M$_{\odot}$}
\def\MMsun{M$_{\odot}$}
\def\Lsun{~L$_{\odot}$}

\title[Gas and Star Formation in the Circinus Galaxy]
      {Gas and Star Formation in the Circinus Galaxy}
\author[B.-Q. For, B. S. Koribalski, and T. H. Jarrett]
       {B.-Q. For$^{1,2}$ \thanks{E-mail:biqing.for@uwa.edu.au}, 
        B. S. Koribalski$^{3}$ and T. H. Jarrett$^{4}$\\
$^{1}$Department of Astronomy, University of Texas, Austin, TX, 78712, USA\\
$^{2}$ICRAR, University of Western Australia, Crawley, WA, 6009, Australia\\
$^{3}$Australia Telescope National Facility, CSIRO Astronomy and Space Science, 
      Epping, NSW 1710, Australia\\
$^{4}$Spitzer Science Center, California Institute of Technology, 
      Pasadena, CA, 91125, USA}

\begin{document}

\date{Accepted 2012 May 30.  Received 2012 May 27; in original form 2012 April 2}

\pagerange{\pageref{firstpage}--\pageref{lastpage}} \pubyear{2012}

\maketitle

\label{firstpage}

\begin{abstract}
We present a detailed study of the Circinus Galaxy, investigating its star 
formation, dust and gas properties both in the inner and outer disk. To 
achieve this, we obtained high-resolution {\it Spitzer} mid-infrared images 
with the IRAC (3.6, 5.8, 4.5, 8.0~$\mu$m) and MIPS (24 and 70~$\mu$m) 
instruments and sensitive \HI\ data from the Australia Telescope Compact 
Array (ATCA) and the 64-m Parkes telescope. These were supplemented by CO 
maps from the Swedish-ESO Submillimetre Telescope (SEST). Because Circinus 
is hidden behind the Galactic Plane, we demonstrate the careful removal of
foreground stars as well as large- and small-scale Galactic emission from 
the {\it Spitzer} images. 
We derive a visual extinction of $A_{\rm V}$ =  2.1~mag from the Spectral 
Energy Distribution of the Circinus Galaxy and total stellar and gas masses 
of $9.5 \times 10^{10}$\Msun\ and $9 \times 10^9$\Msun, respectively. 
Using various wavelength calibrations, we find 
obscured global star formation rates between 3 and 8\Msun\,yr$^{-1}$. Star 
forming regions in the inner spiral arms of Circinus, which are rich in \HI, 
are beautifully unveiled in the {\it Spitzer} 8~$\mu$m image. The latter is 
dominated by polycyclic aromatic hydrocarbon (PAH) emission from heated 
interstellar dust. We find a good correlation between the 8~$\mu$m emission 
in the arms and regions of dense \HI\ gas. The (PAH 8~$\mu$m) / 24~$\mu$m 
surface brightness ratio shows significant variations across the disk of 
Circinus.
\end{abstract}


\section{Introduction\label{intro}}

The Circinus Galaxy (HIPASS J1413--65), or Circinus in short, is a nearby, 
highly obscured spiral galaxy which was discovered by \citet{Freeman77}. 
It is located behind the Galactic Plane ($b \sim -4$\degr), hidden behind a 
veil of dust and high star density. \citet{Freeman77} estimated a Holmberg 
radius of $\sim$17\arcmin, a morphological type of Sb to Sd and a distance 
of around 4~Mpc. 

Circinus hosts an active galactic nucleus (AGN), superluminous H$_2$O masers,
a recent supernova, a star-forming nuclear ring, a spectacular pair of radio 
lobes and a gigantic hydrogen disk. In the following we briefly describe these 
and other features before focusing on the inner disk of Circinus where {\it 
Spitzer} data reveal the stellar disk and the onset of two symmetric spiral 
arms as never seen before.

\citet{Matt96} show that the X-ray spectrum of Circinus is consistent with 
Compton scattering and fluorescent emission from cold matter, illuminated by 
an obscured AGN. Strong far-infrared emission detected from the nuclear 
region suggests high starburst activity \citep{Ghosh92}. Subsequent spectral 
line observations by \citet{Oliva94,Oliva98} confirmed a Seyfert\,2 nucleus.
The detection of a circum-nuclear ring in the H$\alpha$ line is reported by 
\citet{Marconi94} (see also \citealp{Koribalski96}; \citealp{Elmouttie98}). 
Circinus also hosts highly variable H$_{2}$O megamasers \citep{GW82,WG86} 
which appear to arise in the circumnuclear accretion disk and in a 
low-velocity outflow \citep{McCallum09}. \citet{Bauer01} discovered a recent 
supernova, designated SN1996cr and located just 30\arcsec\ south of the 
nucleus of Circinus, and present a multi-wavelength analysis of its 
lightcurve. High-resolution Australia Telescope Compact Array (ATCA) radio 
continuum images reveal a prominent pair of linearly polarized radio lobes 
orthogonal to the disk (\citealp{Elmouttie98,Wilson11}). 

Low-resolution 21-cm observations with the 64-m Parkes radio telescope reveal 
the enormous \HI\ envelope of the Circinus Galaxy. Its large-scale gas disk 
extends over an area more than one degree in diameter and is much larger than 
the stellar disk (\citealp{Mebold76,Freeman77,Henning00,K2004}). 
ATCA \HI\ observations show in detail the distribution and kinematics of the neutral 
gas. Circinus has a very extended and significantly warped \HI\ disk with a 
strong but irregular spiral pattern and an inner bar \citep{Jones99,Curran08}. 
Its \HI\ mass, \MHI, is $7 \times 10^{9}$\Msun\ 
(\citealp{Freeman77,Henning00,K2004}) and its \HI\ diameter is around 
100~kpc. It has a Tully-Fisher estimated distance of 4.2~Mpc \citep{Tully09}.
For the analysis to follow, we adopt a distance of $D$ = 4.2~Mpc.

The \HI\ physical diameter of the Circinus Galaxy is similar in size to that of the 
M\,83 galaxy \citep{K08}. While Circinus is relatively isolated 
and mostly undisturbed, M\,83 is surrounded by numerous dwarf galaxies, e.g.
NGC~5253 \citep{LS2012}, and clearly affected by its environment. The tidally
distorted outer \HI\ arm and associated debris (close to NGC~5264) as well 
as a faint stellar stream provide evidence for gas accretion and weak 
gravitational interactions (Koribalski et al. 2012, in prep.). The large 
\HI\ disks of Circinus and M\,83, which are typical for their \HI\ mass \citep{BvW94}, 
allow us to measure their dark matter content out to radii of $\sim$50~kpc. 
High-resolution GALEX images of M\,83 reveal an extended UV- or XUV-disk, 
suggesting that star formation is not restricted to the inner stellar disk 
\citep{Thilker05}. High-density \HI\ emission is an excellent tracer of
star formation in the outer disk of galaxies; good examples are NGC~1512 
\citep{BA09},  M\,83 (Koribalski et al. 2012, in prep.) and Circinus (this 
paper). 

Despite the richness of interesting features in the Circinus Galaxy, most
studies have been limited to its nuclear region. While the 2MASS Large Galaxy 
Atlas \citep{Jarrett03} has provided an extensive view of the stellar content 
of Circinus, the formidable dust opacity at 2~$\mu$m and the dense stellar 
foreground prevent a detailed investigation of its large scale and low surface 
brightness disk. The mid-infrared (MIR) window is ideal to unveil both the 
old stellar disk population of the obscured Circinus Galaxy as well as the 
interstellar emission from the new born stars in the outer disk. The high 
angular resolution (2\arcsec\ in the IRAC 8~$\mu$m band) and sensitivity of 
the {\it Spitzer} telescope makes it more capable than previous infrared 
observatories (e.g., IRAS, ISO \& AKARI). The on-board {\it Spitzer} high dynamic 
range IRAC and MIPS imaging instruments allow us to study the full stellar 
extent of the Circinus disk and the distribution of its warm and cold dust. 

The IRAC 3.6 and 4.5~$\mu$m bands primarily trace the old stellar component 
in a galaxy while the 5.8 and 8.0~$\mu$m bands reveal polycyclic aromatic 
hydrocarbon (PAH) emission from the heated interstellar dust via UV photons 
radiating from massive young stars. The MIPS 24~$\mu$m band is dominated 
by thermal emission from warm ($T > 120$~K) dust inhabiting the molecular and
neutral medium phases of the ISM, while the 70~$\mu$m band traces relatively
cold ($T < 50$~K) emitting dust. The relative contribution of the hot dust 
emission depends on the strength of the radiation field. It has been shown by 
many semi-empirical dust emission models that an increased radiation field 
leads to a more rapid increase of the luminosity in the 24~$\mu$m band than 
other IR bands (\citealp{LD01}; \citealp{DL07}).  

In this paper, we present a detailed study of the Circinus Galaxy, in
particular, we focus on its star-forming disk and inner spiral arms. This 
is the first analysis of Circinus using {\it Spitzer} images. We compare 
the mid-infrared images with the gas density of the extended \HI\ disk
to unveil star formation sites in the outer disk of Circinus. 

The observations and data reduction are presented in Sections~2 and 3. 
Sections~4--6 describe the derivation of basic parameters via photometry. 
We derive the stellar and gas masses of Circinus in Section~7. 
Identification of local star formation sites and global star formation rates
using a range of wavelength calibrations are given in Section~8. The relation 
between PAH 8~$\mu$m emission and 24~$\mu$m hot dust emission is explored
in Section 9. Finally, a summary and conclusions are given in Section~10. 

\section[]{Observations\label{obs}}

\subsection{{\it Spitzer} Observations\label{spitzerobs}}

The mid-infrared imaging observations of the Circinus Galaxy were carried 
out with the IRAC (3.6, 4.5, 5.8, and 8.0~$\mu$m) and MIPS (24 and 70~$\mu$m) 
bands of {\it Spitzer} under the program led by T. Jarrett (PID 50173). The 
objective was to survey a region large enough to cover the entire \HI\ disk 
of Circinus, extending well beyond a smaller {\it Spitzer}  survey of the disk 
(G. Rieke, PID 40936),
with a sensitivity to capture the low surface brightness emission 
in the extended star-forming disk.  

For both {\it Spitzer} instruments, two 
separate Astronomical Observing Request (AOR) observations (epochs) were 
combined to improve the sensitivity and to mitigate data artifacts and 
radiation events.  

The IRAC observations consisted of a series of 12~s exposures in high dynamic 
range mode (HDR) using a $20 \times 19$ grid with 154\arcsec\ steps (roughly 
50\% overlap). Including both epochs, the total field coverage is $50\arcmin
\times 50\arcmin$ with a pixel scale of 0\farcs6. Each pixel has an exposure 
depth of roughly two epochs $\times$ 12~s $\times$ four coverages, i.e., 96~s 
per pixel. The achieved 1$\sigma$ sensitivity in surface brightness is 0.059, 
0.045, 0.073 and 0.086 MJy\,sr$^{-1}$, respectively, for the 3.6, 4.5, 5.8 
and 8.0~$\mu$m bands. The short-exposure (0.6~s) HDR observations were used 
to recover the saturated nucleus of Circinus.

The MIPS observations consisted of mapping using the medium scan rate with 
21 total scan legs each separated by 148\arcsec. Including both epochs, the 
total area covered is $\sim$0.9\degr $\times$ 0.9\degr. Each pixel has a 
depth of coverage or redundancy of roughly 50. The achieved 1$\sigma$ 
sensitivity in surface brightness is 0.058 and 1.46 MJy\,sr$^{-1}$, 
respectively, for the 24 and 70~$\mu$m bands.  

We used tools developed by the Spitzer Science Center (SSC) to identify and 
correct the basic calibrated data (BCD) frame pixels potentially affected by 
muxbleed (arising from bright stars in the field). After BCD image correction, 
we used the SSC-developed software package MOPEX to perform background 
matching, temporal and spatial outlier rejection, astrometric pointing 
refinement, and co-addition of BCDs into large mosaics for the IRAC and 
MIPS 24~$\mu$m observations. In addition to the co-added signal mosaic, the 
MOPEX 
products include the uncertainty and coverage maps. For the MIPS 70~$\mu$m 
observations, instead of using MOPEX to produce our own mosaics, we combined 
both epochs of the SSC-produced post-BCD (PBD) mosaics, which have adequate 
quality for science analysis.

\begin{figure*}
\center
\includegraphics[width=17cm]{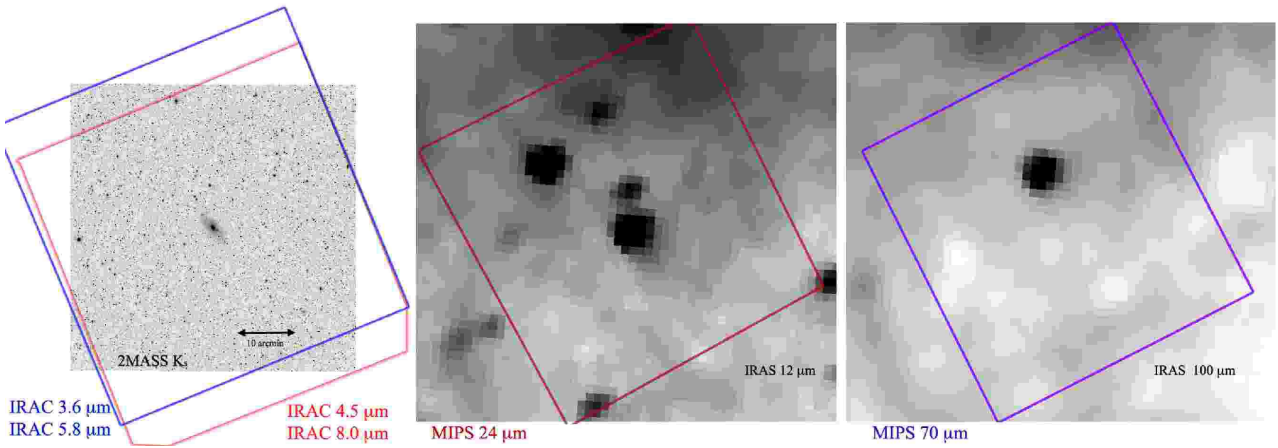}
\caption{{\it Spitzer} AOR mapping footprints.  The first panel shows the 
IRAC footprints overlayed on a 2MASS K$_s$-band image.
The 2nd panel shows the MIPS-24 map, overlayed on the
IRAS 12~$\mu$m  image, and the third panel shows the MIPS-70
map overlaying the IRAS 100~$\mu$m  image.  The field of view for
each panel is about is $1\degr 
   \times 1\degr$, with North pointing up and East pointing to the left.}
\label{AORs} 
\end{figure*}

The achieved angular resolutions are 2\arcsec\ for all IRAC bands, 6\arcsec\ 
for the MIPS 24~$\mu$m band and 18\arcsec\ for the MIPS 70~$\mu$m band. We 
summarize the observing parameters in Table~\ref{obspar} and show the coverage 
footprints of IRAC and MIPS channels in Figure~\ref{AORs}. 

\begin{table*} 
\caption{{\it Spitzer} Observing Parameters for the Circinus Galaxy.
  \label{obspar}}
\begin{tabular}{lcc}
\hline
\hline
observing dates and pointing positions: 
 & \multicolumn{2}{r}{20 March 2009,
   $14^{\rm h}\,13^{\rm m}\,10^{\rm s}$, --65\degr\,20\arcmin\,21\arcsec} \\
 & \multicolumn{2}{r}{26--27 March 2009,
   $14^{\rm h}\,13^{\rm m}\,14^{\rm s}$, --65\degr\,21\arcmin\,30\arcsec} \\
\hline
               & IRAC                         & MIPS \\
\hline
coverage area  & $50\arcmin \times 50\arcmin$ & $54\arcmin \times 52\arcmin$\\
grid size      & $19 \times 20$               & 21 scan legs \\
spacecraft orientation & --69\degr            & +30\degr \\
bands ($\mu$m) & 3.6, 4.5, 5.8, 8.0           & 24, 70 \\
total exposure time per pixel & 96\,s         & $\sim$50 depth \\
$1\sigma$ surface brightness limits  (MJy\,sr$^{-1}$) 
               & 0.059, 0.045, 0.073, 0.086   &   0.058, 1.46 \\
\hline
\end{tabular}
\end{table*}

\subsection{ATCA \HI\ Observations\label{atcaobs}}

High-resolution \HI\ observations of the Circinus Galaxy were carried out
with the ATCA, located in Narrabri, Australia. Three array configurations 
(375, 750A and 1.5) were used to obtain single pointings. For details see 
\citet{Jones99}, who analyze the large-scale \HI\ distribution and kinematics 
of Circinus at an angular resolution of about 1$\arcmin$ (rms noise $\sim$1.3 
mJy\,beam$^{-1}$ per 6.6\kms\ channel). To study the full \HI\ extent of 
Circinus, which is much larger than the 33$\arcmin$ ATCA primary beam, \HI\ 
mosaic observations were obtained in the 375-m array. For details see 
\citet{Curran08} who analysed both the large-scale \HI\ emission of Circinus 
(at a resolution of $124\arcsec \times 107\arcsec \times 4$\kms) as well as 
the inner CO disk. The latter was observed  with the SEST 15-m dish at a 
resolution of at 45\arcsec\ for CO(1--0) and at 22\arcsec\ for CO(2--1) and 
1.8\kms\ channel width. 

As Circinus lies in the Zone of Avoidance, Galactic \HI\ emission is prominent 
and detected at radial velocities between about --280 and +50\kms. \HI\ 
emission from Circinus is observed between +240 and +640\kms. Its \HI\ 
diameter is at least a factor of five larger than the largest estimated 
optical and infrared diameters.

For the data analysis in this paper we created (1) Galactic \HI\ maps and
(2) high-resolution \HI\ maps of the inner disk of the Circinus Galaxy for
comparison with the {\it Spitzer} MIR data. For both tasks we used the
{\sc miriad} software package \citep{Sault95}.

For the detailed comparison of the Spitzer mid-infrared data with the \HI\ 
data, we need to substantially improve the angular resolution. This can be
achieved by robust by robust or uniform weighting of the $uv$-data, i.e. 
giving more weight to the longest baselines in the array ($\la$6~km). We use 
the calibrated ATCA $uv$-data from \citet{Jones99} plus \HI\ data taken in 
the 6A and 6C array configurations to make \HI\ cubes of the inner Circinus 
disk at an angular resolution of 15\arcsec. At this resolution, achieved 
with robust = 0 weighting, good signal-to-noise and high image fidelity was 
obtained.

To estimate the foreground emission, most prominent in the {\it Spitzer} IRAC 
8~$\mu$m and MIPS 24~$\mu$m images, we obtained Galactic \HI\ maps of the 
area using $uv$-data from (a) the high-resolution single-pointing data by 
\citet{Jones99} and (b) the low-resolution ATCA mosaic by \citet{Curran08}. 
The latter was combined with the stray-radiation corrected \HI\ data from 
the Galactic All-Sky Survey (GASS; \citealp{MG09}; \citealp{Kalberla10}) to 
estimate the large-scale diffuse Galactic \HI\ emission. After successful 
subtraction of the scaled Galactic \HI\ map from the {\it Spitzer} images, 
some small-scale features remain. Investigation of the Galactic \HI\ emission 
at high resolution revealed a clear correlation between the small-scale 
structure and Galactic \HI\ emission over a specific velocity range (see
Section~\ref{MWsub} for details). 

\section{Data Reduction\label{reduction}}

\subsection{{\it Spitzer} Images\label{spitzerred}}

\begin{figure*}
\center
\includegraphics[width=5.5cm]{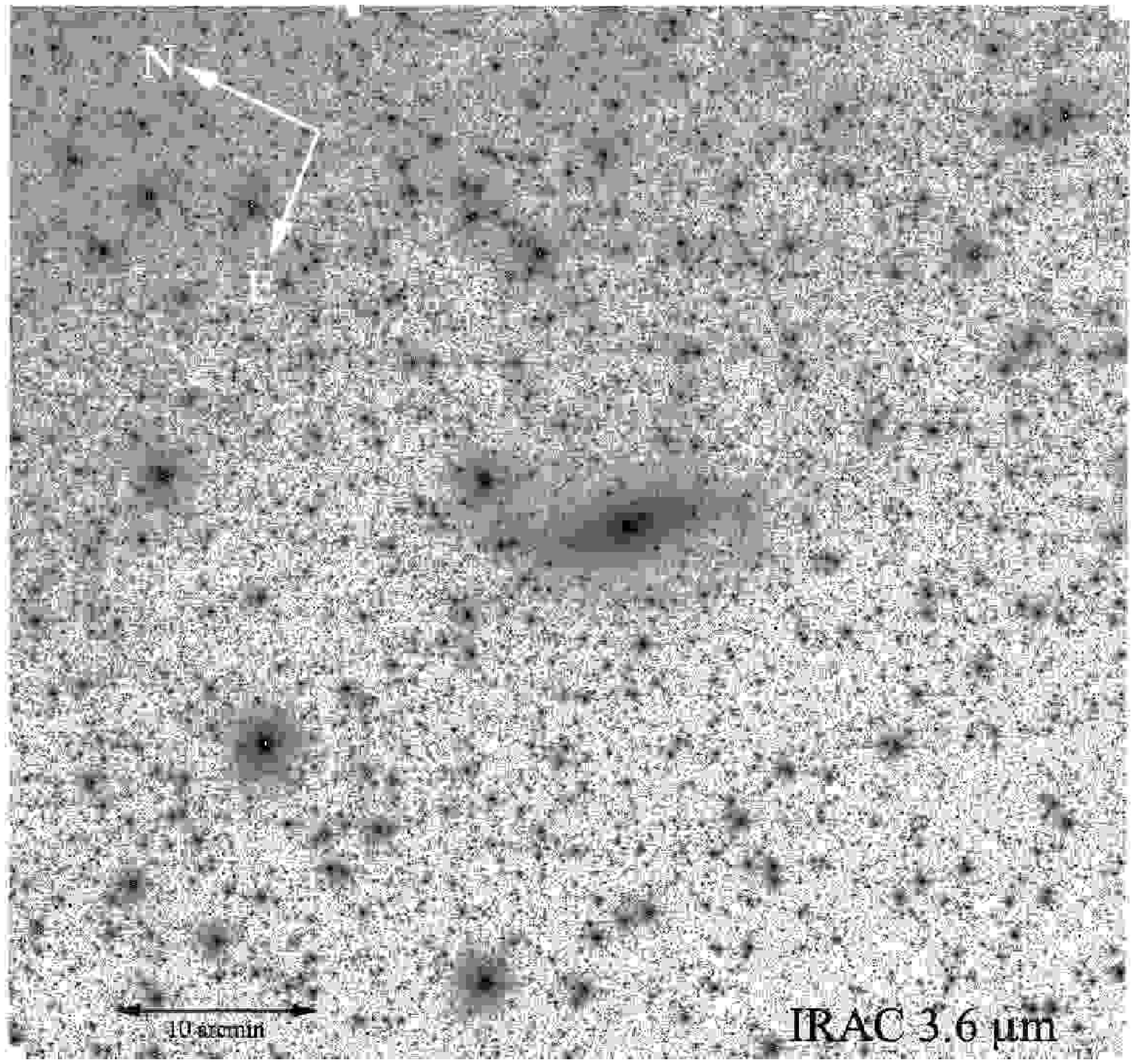}
\includegraphics[width=5.5cm]{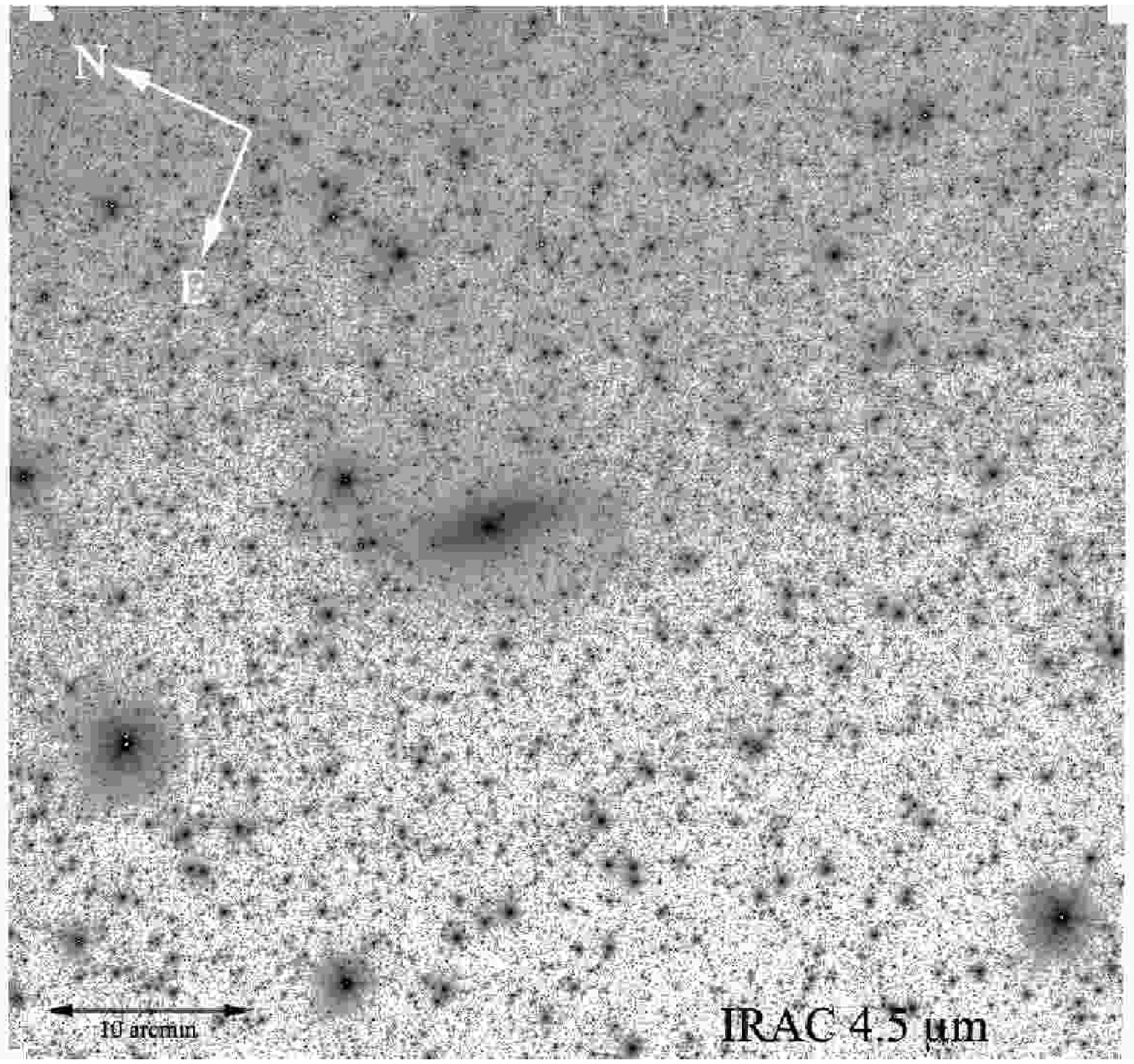}
\includegraphics[width=5.5cm]{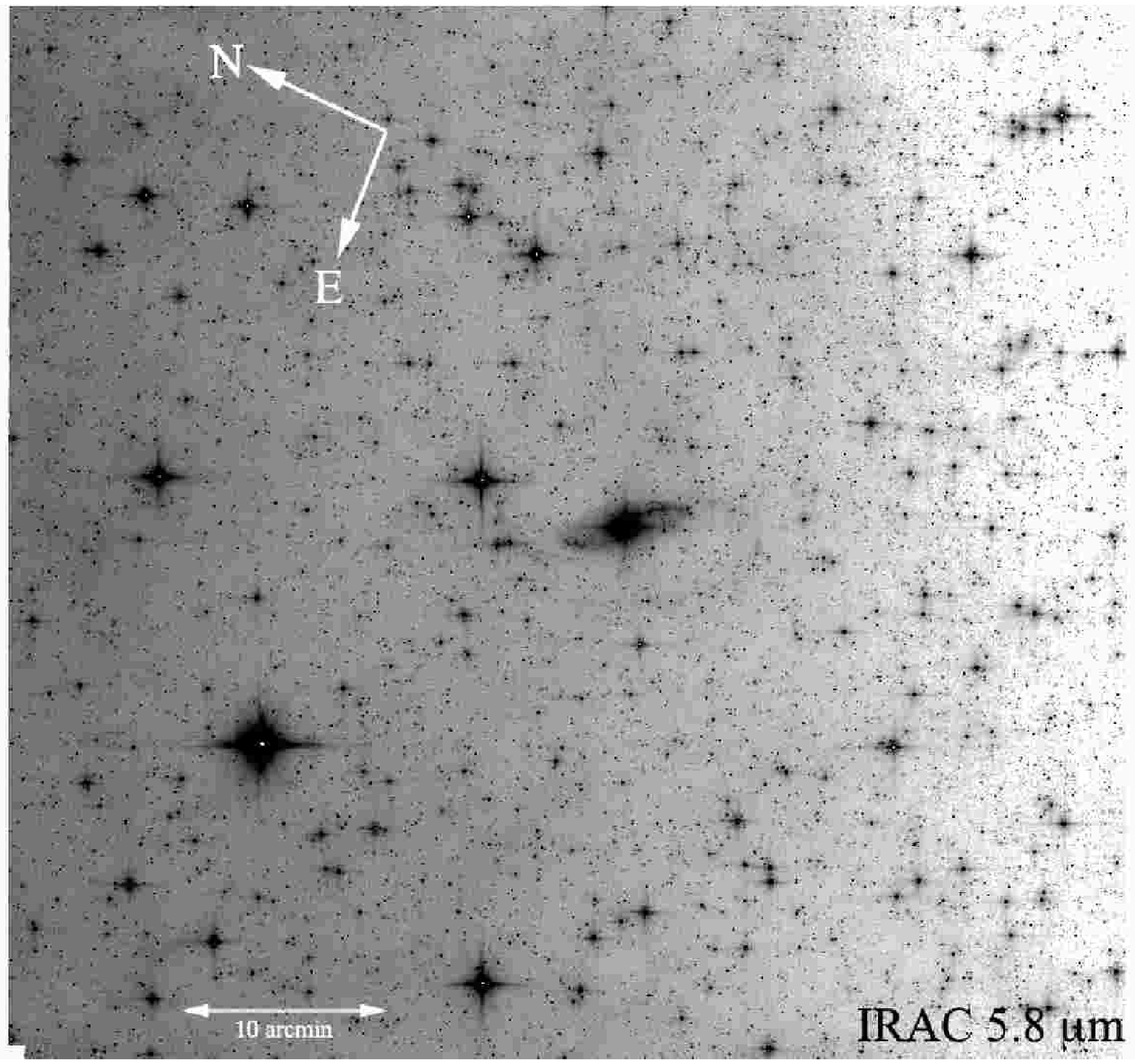}
\includegraphics[width=5.5cm]{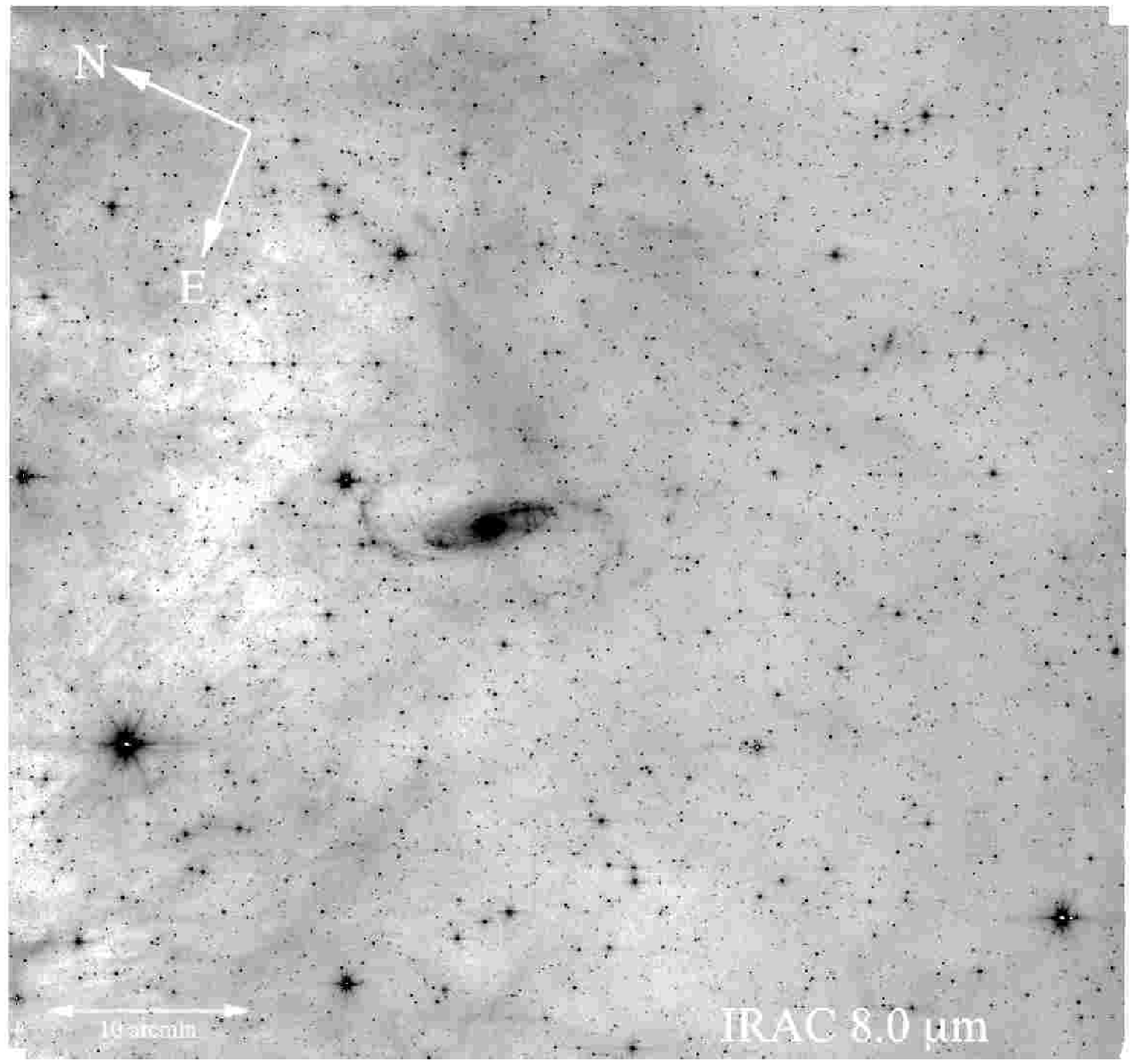}
\includegraphics[width=5.5cm]{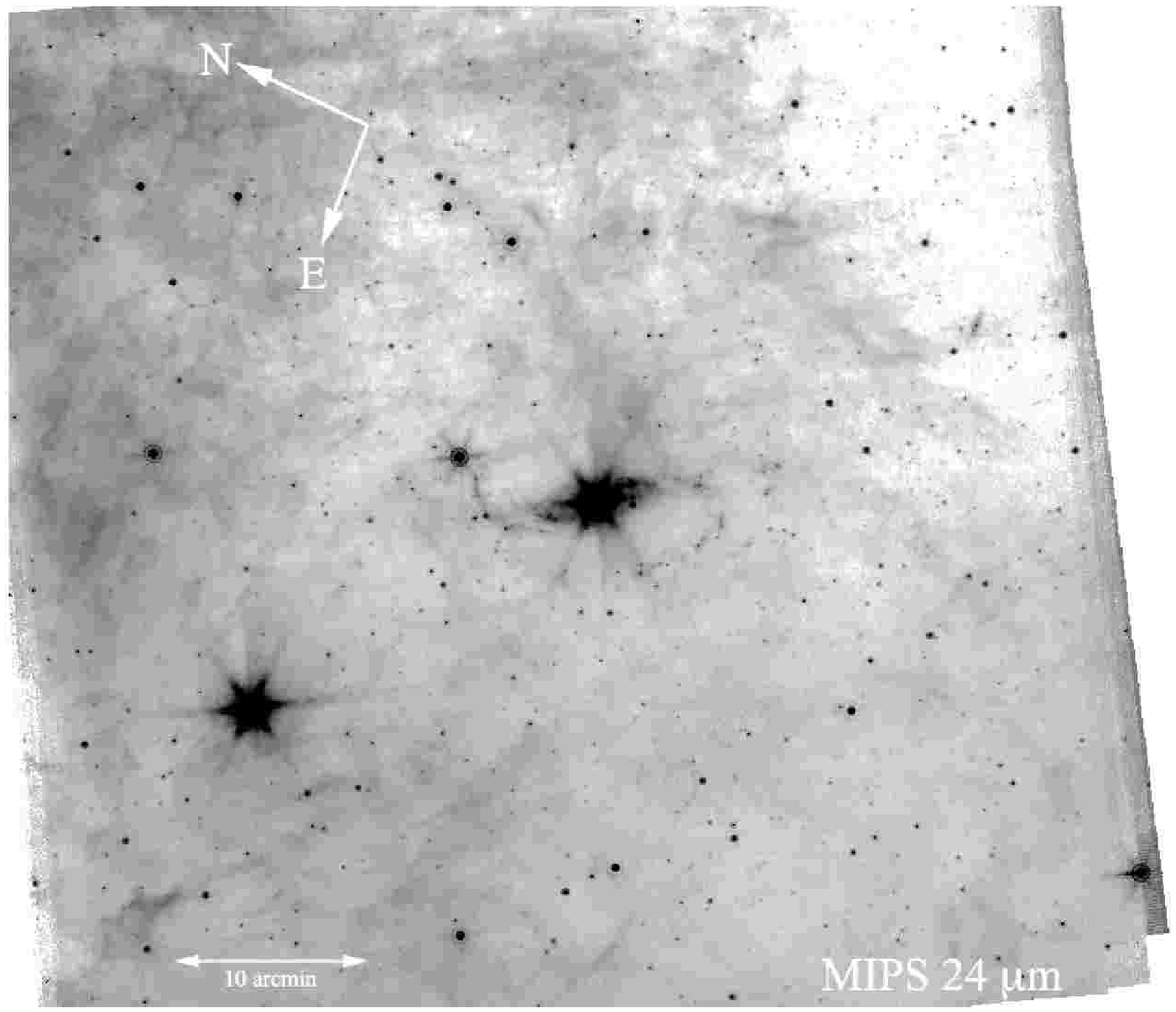}
\includegraphics[width=5.5cm]{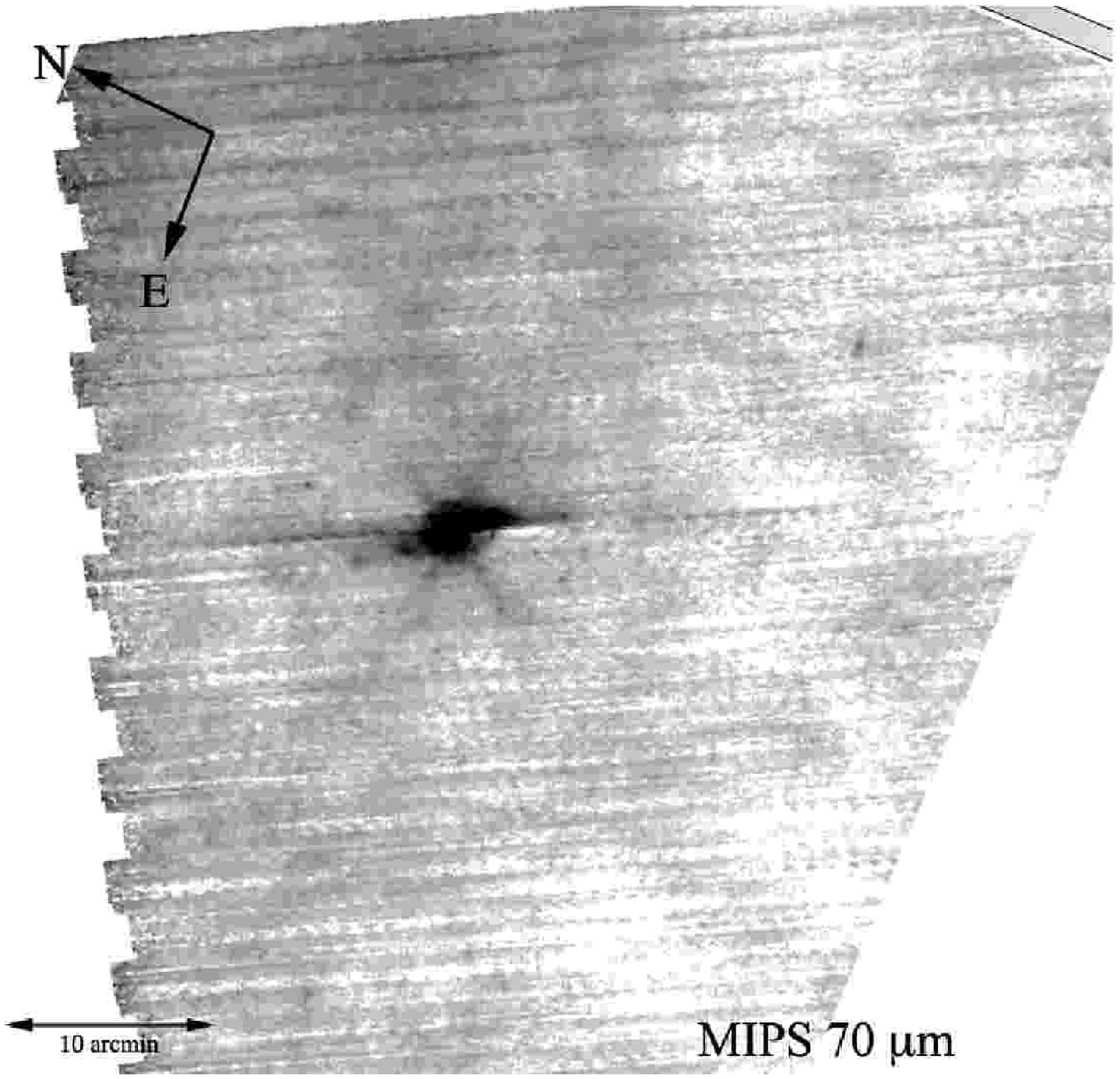}
\caption{{\it Spitzer} 3.4 to 70~$\mu$m images of the Circinus Galaxy. The 
   observed field of view is nearly one degree across. The image orientation 
   corresponds to the original spacecraft rotation angle at the time of the 
   observations. The North and East directions are indicated by arrows, and
   a 10\arcmin\ scale bar is given at the bottom left.}
\label{bandscomp}
\end{figure*}

The location of the Circinus Galaxy, obscured by the Galactic Plane, presents 
a challenge for source characterization. In order to determine its global 
properties, including its shape, size and photometric properties, as well as 
study its extended low surface brightness star-forming disk, the foreground 
Milky Way (MW) stars and Galactic ISM emission must first be identified and 
subtracted from the {\it Spitzer} images. Figure~\ref{bandscomp} shows the 
reduced IRAC and MIPS images with the Circinus Galaxy at the center, obscured 
by thousands of stars and wide-spread dust emission from the Galactic Plane.

To identify and subtract MW stars, we used detection and characterization 
tools that were designed for the multi-band imaging observations obtained
with the {\it Spitzer} and WISE space-borne infrared telescopes, developed 
by the WISE Science Data Center (WSDC) of the Infrared Processing and 
Analysis Center (IPAC). The distinguishing characteristic of these systems 
is that they use all bands simultaneously to detect and characterize sources.
The detection system is described in \citet{MJ11} and the source 
characterization system in Jarrett et al. (2012) and \citet{Cutri11}. For the 
Circinus observations, we detect $\sim$90,000 sources, dominated by the IRAC 
3.6 and 4.5~$\mu$m sensitivity to Galactic stars. Not all of these detections
are of Galactic origin, many are indeed associated with the Circinus Galaxy. 
In order to mitigate removal of real emission associated with Circinus, 
we visually inspected the images, taking special notice of detected sources 
that were either in close proximity to Circinus spiral arms, or were 
spatially resolved, or had very red colors: those with colors or spatial 
properties that were more consistent with association to Circinus rather with 
the Galactic foreground population, were not removed from the images. Using 
the WISE source characterization system, the surviving Galactic population 
was then subtracted from the images using the individual IRAC or MIPS point 
spread function (PSF) to fit the stellar profiles. The resulting images have 
their MW stellar component removed.

\subsection{Galactic ISM Subtraction\label{MWsub}}

The thermal dust emission in the near-IR is caused by a mixture of large and 
small grains, which dominate the total dust mass of a galaxy. Here we compare
the spatial distribution of dust emission, derived from {\it Spitzer} MIR
maps, to that of the \HI\ gas. By conducting a correlation analysis, we aim
to separate Galactic from extragalactic (Circinus Galaxy) infrared emission.
Specifically, we use single dish and interferometric 21-cm \HI\ data towards
the Circinus Galaxy to correct for the Galactic foreground emission present 
in IRAC 8~$\mu$m and MIPS 24~$\mu$m images  (note that the foreground ISM
contribution is negligible in the short-wavelength bands of IRAC). 

We employed the ATCA \HI\ data sets of the Circinus Galaxy described by 
\citet{Jones99} and \citet{Curran08} (see Section~\ref{atcaobs}) as well as
Parkes \HI\ maps from the Galactic All-Sky Survey (GASS). The velocity range 
and resolution of these \HI\ data sets allow a clear separation and analysis 
of both the Galactic features and the extended structure of the Circinus 
Galaxy. 

The high-resolution {\it Spitzer} images (see Figure~\ref{bandscomp}) show 
structures on all scales. While the IRAC 3.6, 4.5, and 5.8~$\mu$m images are 
dominated by scores of foreground stars, and diffuse foreground emission 
(filaments, bubbles, sheets, etc.) is pervasive in the IRAC 8~$\mu$m and 
MIPS 24~$\mu$m images. By combining the Parkes and ATCA \HI\ data towards 
Circinus over the Galactic velocity range, we constructed a high angular 
resolution image that matches the large-scale MIR features. 

\subsubsection{Subtracting the Large-scale Galactic Foreground \label{MWsub1}}
We made ATCA \HI\ mosaic images of the Galactic \HI\ emission (over the 
velocity range from --220 to +220\kms) by Fourier transforming the data 
obtained by \citet{Curran08} supplemented by \HI\ data taken with the very
compact H75 array. The resulting ATCA data cube has a velocity resolution 
of 4\kms\ and an angular resolution of 1\arcmin.

The Parkes \HI\ data were retrieved from the GASS data 2 release 
website\footnote{http://www.astro.uni-bonn.de/hisurvey/gass} and is 
corrected for stray radiation and radio frequency interference
\citep{Kalberla10}. 
GASS covers the region south of $\delta$ = +1\degr\ in the Local Standard 
of Rest (LSR) velocity range from --400 to +500\kms, with 1~\kms\ velocity 
resolution and an angular resolution of 16\arcmin. 

To combine the interferometer (ATCA) and single-dish (Parkes) data, we explore
two methods: (a) merging in the Fourier domain (linear method) and (b) merging 
during deconvolution (non-linear method) (see \citealp{SS02}). For both 
methods, the GASS data were converted from Kelvin to Jy\,beam$^{-1}$, and 
from LSR to the barycentric (heliocentric) velocity reference frame, then 
regridded to match the ATCA mosaic pixel scale and velocity channels. We 
applied the residual primary beam attenuation (a gain function) present in 
the ATCA mosaic to the GASS data prior to combining them.

\begin{figure*}
\center
\includegraphics[scale=0.6]{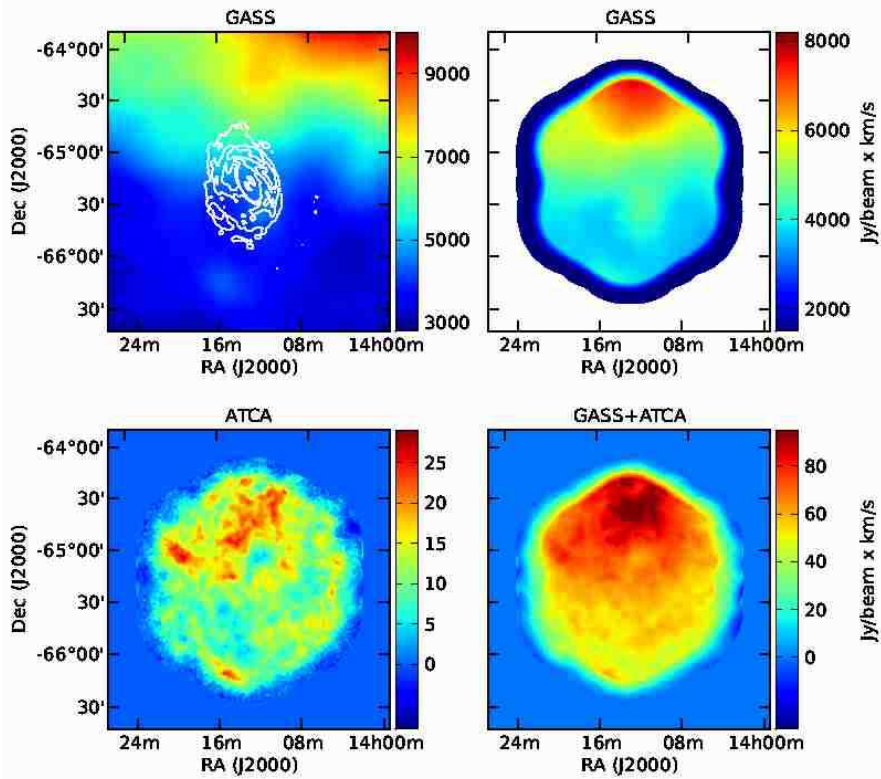}
\caption{Galactic \HI\ emission in the direction of the Circinus Galaxy. 
   {\bf Top}: We show the integrated Galactic \HI\ distribution as obtained 
     from the Parkes \HI\ data (left) and tapered with the gain function of
     the ATCA \HI\ mosaic (right). For comparison we overlay the large-scale 
     \HI\ distribution (white contours) of the Circinus Galaxy; the contour 
     levels are 0.5, 3.0 and 6.0~Jy\,beam$^{-1}$\kms.
   {\bf Bottom}: We show the integrated Galactic \HI\ distribution as obtained 
     from the ATCA \HI\ mosaic observations (left) and the combined Parkes
     and ATCA \HI\ data set (right). The latter is used to remove the 
     large-scale Galactic foreground emission from the {\it Spitzer} IRAC 
     8~$\mu$m and MIPS 24~$\mu$m images.}
\label{moment0}
\end{figure*}

To generate the residual image, a deconvolved and restored ATCA mosaic is 
needed. We first applied the mosaic Steer CLEAN algorithm \citep{SDI84} to 
deconvolve the image, but ultimately failed due to its limitation in handling diffuse 
emission. In its place, we employed a maximum-entropy based deconvolution algorithm 
({\sc mosmem}), which we found to work better on diffuse emission. The only 
drawback is that {\sc mosmem} does not handle negative features. 
Subsequently, an integrated \HI\ map was created using all channels with 
Galactic \HI\ emission (velocities ranging from --88 to +124\kms).
The rms achieved is around 4.5~mJy\,beam$^{-1}$.  
Around 93\% 
of the total \HI\ flux was recovered in the combined Galactic map as 
compared to the single-dish data alone.  
Figure~\ref{moment0} illustrates the transformation of the individual maps to
the combined result.

The procedure for subtracting the Galactic foreground from the {\it Spitzer} 
images was as follows: 
---(a) A constant was added to the calibrated IRAC 8~$\mu$m and MIPS 24~$\mu$m 
    images to account for some over-subtraction of the background during 
    mosaic construction. The offset was determined based on the post-BCD 
    mosaics delivered by SSC, where the background had not been subtracted. 
---(b) The resulting {\it Spitzer} images were then divided by the integrated 
    Galactic \HI\ emission map to determine the respective correlation factors. 
---(c) We then subtracted the scaled integrated Galactic \HI\ emission maps
    from the {\it Spitzer} images produced in step (a).
---(d) A residual smooth gradient over the full IRAC 5.8~$\mu$m and 8~$\mu$m images, known as 
    the ``first frame effect'' (IRAC Instrument Handbook) was fitted and removed using a 
    first order polynomial (after masking the Circinus Galaxy and another 
    background galaxy). 

\begin{figure*}
\includegraphics[width=17cm]{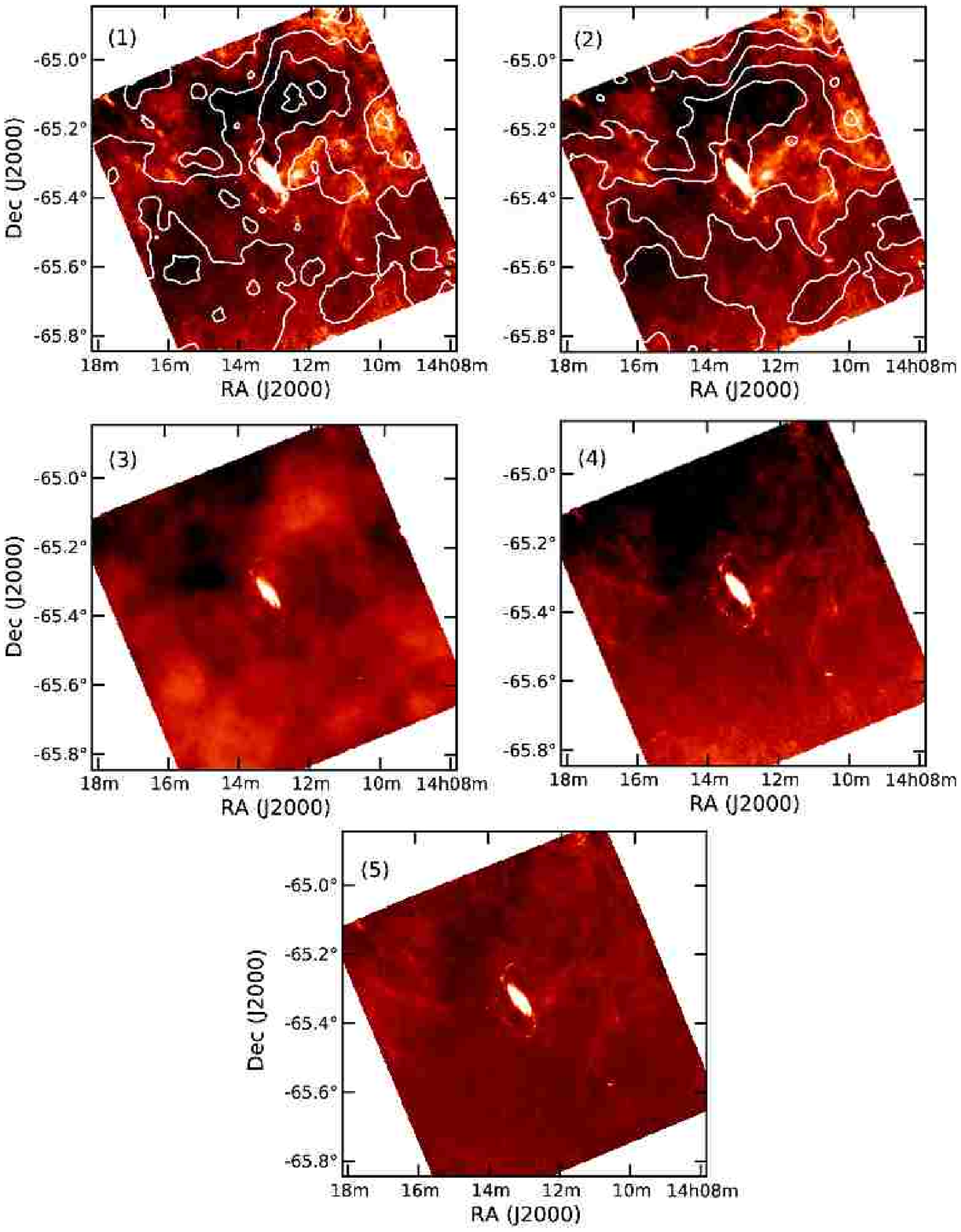}
\caption{{\it Spitzer} IRAC 8~$\mu$m images of the Circinus Galaxy and 
   surroundings after the removal of foreground stars and Galactic emission. 
   White contours indicate the Galactic \HI\ emission as 
   obtained from the ATCA mosaic observations (panel 1) and the combined 
   Parkes and ATCA data (panel 2). The results of subtracting the scaled 
   Galactic \HI\ emission from the 8~$\mu$m image are shown in panels (3) 
   and (4), respectively. After removing the smooth gradient from the panel 4 result, 
   we obtain a much cleaner IRAC 8~$\mu$m image of Circinus (panel 5).}
\label{foresub_proc} 
\end{figure*}

In Figure~\ref{foresub_proc} we show the IRAC 8~$\mu$m image before and after 
foreground subtraction. The \HI\ contours trace a range of structures, much 
of which correlate well with the infrared dust emission. In particular, the 
Parkes \HI\ data combined with the ATCA \HI\ mosaic allow us to match and 
subtract the diffuse foreground emission on large scales. Some small-scale 
structure remains, most noticeably the ``spur'' structure to the west and overlaying the southwest disk of
Circinus.  It was recently
pointed out that the MIR PAH emission of M\,83, associated with current star formation,
was highly correlated with dense (n $>$ 10$^{2}$ cm$^{-3}$) \HI\ emission (Jarrett et al. 2012).  
It likely 
that the dense filamentary components of the Galactic ISM will also be 
correlated with the MIR PAH emission.  Isolating the dense \HI\ gas  is the
key to identifying and removing the ``spur" contaminant from the {\it Spitzer} imaging. 
The removal of this filament is described in the next subsection.

\subsection{The Galactic Spur\label{spur}}

We re-analyzed single pointing 21-cm data of the Circinus Galaxy taken by 
\citet{Jones99} with three configurations of the ATCA (see Section~\ref{atcaobs}). 
In addition to the detailed \HI\ studies of Circinus itself, these data 
allowed us to make high-resolution maps of the Galactic \HI\ emission. Our 
aims are (a) to map the Galactic \HI\ emission and search for features 
matching PAH structures as seen in the {\it Spitzer} IRAC 8~$\mu$m and MIPS 
24~$\mu$m images and (b) to map the distribution and dynamics of the 
high density \HI\ gas in the inner disk of the Circinus Galaxy.

\begin{figure*}
\begin{tabular}{cc}
   \includegraphics[width=17cm]{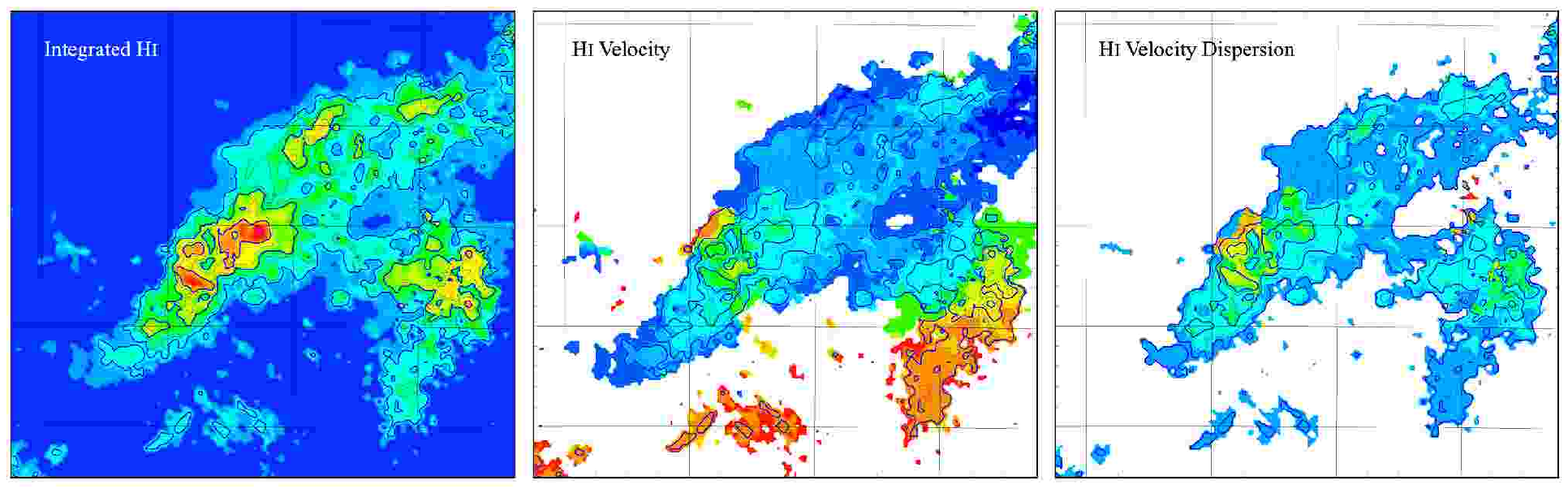}  
\end{tabular}
\caption{High resolution ATCA \HI\ integrated and velocity distribution
of the Galactic foreground to the Circinus Galaxy.  
The velocity ranges between --50 to --30\kms. 
{\bf Left}: Integrated \HI\ distribution (0th 
   moment). 
   {\bf Middle}: Mean \HI\ velocity field (1st. moment). The 
   main feature (blue) has velocities around --45\kms, while the 
   small SW cloud (orange) has velocities around --35\kms. 
   {\bf Right}: Mean \HI\ velocity 
   dispersion (2nd. moment) with maximum velocities around 7\kms.}
\label{HIspur2}
\end{figure*}

\begin{figure*}
\begin{tabular}{cc}
  \includegraphics[width=17cm]{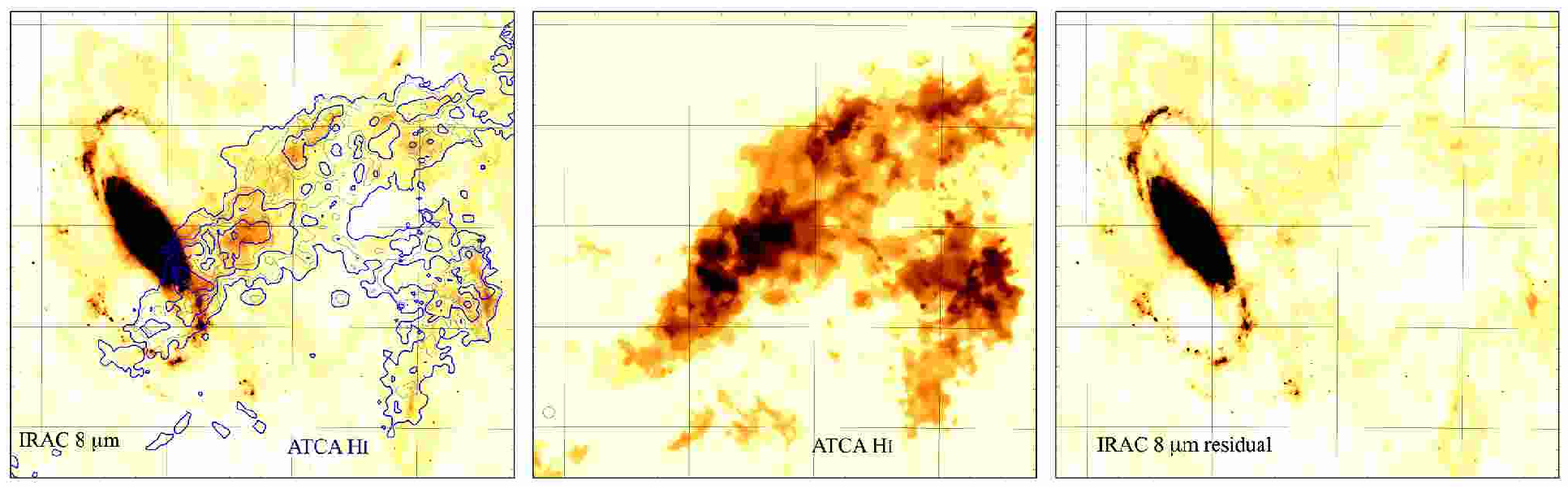} 
\end{tabular}
\caption{ {\it Spitzer} IRAC 8~$\mu$m imaging of the Circinus Galaxy and the 
   Galactic foreground as traced in ATCA \HI\ maps at high angular resolution. 
   {\bf Left}: IRAC 8~$\mu$m image with 
   large-scale Galactic emission removed (for details see Section~\ref{MWsub1}),
    overlaid with ATCA \HI\ contours of the Galactic spur.
    The contour levels are 0.2, 
     0.3, 0.4, 0.5 and 0.6 Jy\,beam$^{-1}$ \kms; the first contour corresponds 
     to an \HI\ column density of $1.8 \times 10^{20}$~cm$^{-2}$.
   {\bf Middle}: ATCA \HI\ distribution of the Galactic spur. 
   The spur was identified in the
     \HI\ spectral cube, velocities between --50 and --30\kms (see Fig. \ref{HIspur2}).
   The synthesized beam of $\sim$35\arcsec is indicated at the bottom left.
   {\bf Right}: IRAC 8~$\mu$m image with the Galactic spur removed from the
   mid-infrared emission.
   The coordinate grid is in steps of 5\arcmin\ in Dec and 1$^{\rm m}$ in RA.
    }
\label{HIspur1}
\end{figure*}

We Fourier-transformed the $uv$-data from all three configurations to image
the Galactic \HI\ emission over a velocity range from $-$220 to +220\kms\ and 
at a range of angular resolutions. Using robust = 0 weighting, which resulted
in an angular resolution of $\sim$35\arcsec, we were able to identify the 
Galactic spur at velocities ranging from about $-$50 to $-$30\kms. Using the 
primary beam corrected \HI\ data cube, we created moment maps of the feature
(see Figure \ref{HIspur2}). It is a large filamentary structure, spanning diagonally across
the Circinus field, intersecting the southwestern disk of the galaxy.  The bulk of the structure
has a mean velocity of  --45\kms, while further to the southwest, another adjoining gas cloud appears
at a velocity of  --35\kms.
The shape and intensity of the \HI\ spur
is largely correlated with the 8~$\mu$m PAH  (see Figure \ref{HIspur2}), 
which we determine to be a scaling
factor $\sim$0.8.  Accordingly, the \HI\ distribution of 
the spur, scaled by this factor, was subtracted from the IRAC 8~$\mu$m 
image. A reasonable subtraction was achieved as Circinus remained symmetric
and undistorted, thus allowing a much improved view of the 
star-forming inner spiral arms of the Circinus Galaxy. Figure~\ref{HIspur1} 
illustrates the process. We are reasonably confidence that the observed spur is a Galactic  
feature and not associated with the Circinus Galaxy itself. 

Following the analysis of \HI\ emission and absorption line measurements 
towards the pulsar PSR~1401--6357 \citep{Johnston96}, we can assign an 
approximate distance to the Galactic spur. A velocity of --50\kms\ corresponds 
to the tangent point in that direction and a distance of $\sim$5~kpc. The 
integrated \HI\ flux density of the spur is $\sim$140~Jy \kms, corresponding
to an \HI\ mass of $\sim$10$^3$\Msun, and size of $\sim$20\arcmin (30 pc). 

In the remaining analysis, we use the stellar and Galactic-ISM subtracted imaging to measure and
characterize the Circinus galaxy.

\section{Surface Brightness Profiles\label{SBP}}

We obtain the surface photometry of Circinus by fitting elliptical apertures 
to the {\it Spitzer} images in all IRAC bands and the MIPS 24~$\mu$m and 70~$\mu$m bands. 
The measurements were performed for radii from 10\arcsec\ to 600\arcsec, in 
steps of 10\arcsec, assuming a position angle of $PA$ = 210\degr\ and an 
inclination angle of $i$ = 65\degr\ \citep{Curran08}. 

\begin{figure*}
\includegraphics[scale=0.5,angle=-90]{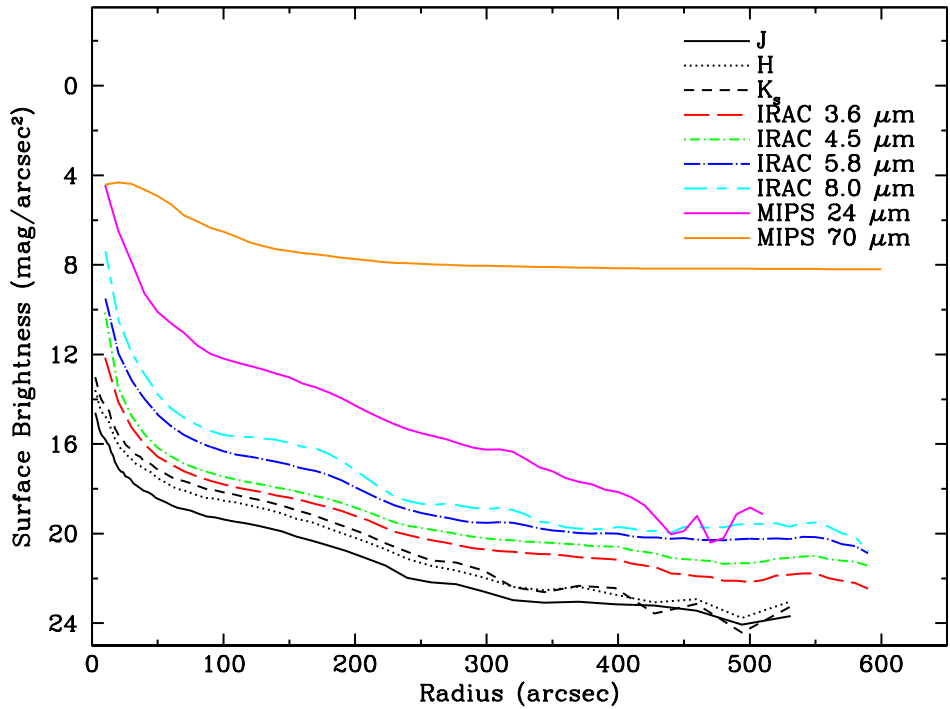}
\caption{Inclination-corrected surface brightness ($cSB$) profiles of the 
   Circinus Galaxy in all four IRAC bands, the MIPS 24~$\mu$m and 70 ~$\mu$m bands, and the 
   2MASS $J$, $H$, and $K_{\rm s}$ bands. See Table~\ref{sbtable} for 
   details; no extinction correction has been applied.}
\label{sb} 
\end{figure*}

The surface brightness ($SB$) profile, in units of mag\,arcsec$^{-2}$, can 
be represented by 
\begin{equation}
  SB = -2.5 \log_{10} (\frac{f_{\rm A} \times F \times C}{F_{0}}), 
\end{equation} 
where $f_{\rm A}$ is an effective aperture correction factor to account for 
extended emission from the IRAC PSF and from the diffuse scattering of the 
emission across the IRAC focal plane, $F$ is the inclination-corrected flux 
density in units of MJy\,sr$^{-1}$, $C$ is a conversion factor, $2.35 \times 
10^{-5}$~Jy\,arcsec$^{-2}$, and $F_{0}$ is the zero-magnitude flux of the 
star Vega. We used $f_{\rm A}$ =  0.944, 0.937, 0.772, 0.737 and $F_{0}$ 
= $280.9 \pm 4.1$, $179.7 \pm 2.6$, $115.0 \pm 1.7$, $64.13 \pm 0.94$ Jy 
for the corresponding IRAC channels 1, 2, 3, 4 from \citet{Reach05}. For 
the MIPS 24~$\mu$m and 70~$\mu$m bands no aperture correction is needed, and we use $F_{0}$
= 7.17~Jy and 0.778~Jy, respectively. 
Table~\ref{sbtable} lists our measured radial flux densities and 
radial surface brightness densities for the Circinus Galaxy. Figure~\ref{sb} 
shows the $SB$ profiles of Circinus as obtained in five {\it Spitzer} bands
as well as the 2MASS $J$, $H$, and $K_{\rm s}$ bands from \citet{Jarrett03}. 

\begin{sidewaystable} 
\tiny
\begin{minipage}{250mm}
\caption{Radial flux density ($F_{\lambda}$) 
   and inclination-corrected radial surface brightness ($cSB$) of the 
   Circinus Galaxy as a function of radius. We used $PA$ = 210\degr\ 
   and $i$ = 65\degr. No extinction correction has been applied.}
\label{sbtable}
\begin{tabular}{r|r|r|c|r|r|c|r|r|c|r|r|c|r|r|c|r|r|c}\\
\hline
\hline
Radius & $F_{\rm 3.6 \mu m}$  & rms & $cSB$ & 
         $F_{\rm 4.5 \mu m}$  & rms & $cSB$ &	
         $F_{\rm 5.8 \mu m}$  & rms & $cSB$ & 	
         $F_{\rm 8.0 \mu m}$  & rms & $cSB$ &
         $F_{\rm 24.0 \mu m}$ & rms & $cSB$ & 
         $F_{\rm 70.0 \mu m}$ & rms & $cSB$ 
\\
(\arcsec) & \multicolumn{2}{|c|}{(MJy\,sr$^{-1}$)}	
          & (mag/(\arcsec)$^{2}$)	
          & \multicolumn{2}{|c|}{(MJy\,sr$^{-1}$)}
          & (mag/(\arcsec)$^{2}$)	
          & \multicolumn{2}{|c|}{(MJy\,sr$^{-1}$)}	
          & (mag/(\arcsec)$^{2}$)	
          & \multicolumn{2}{|c|}{(MJy\,sr$^{-1}$)}
          & (mag/(\arcsec)$^{2}$)	
          & \multicolumn{2}{|c|}{(MJy\,sr$^{-1}$)}
          & (mag/(\arcsec)$^{2}$)	 
          & \multicolumn{2}{|c|}{(MJy\,sr$^{-1}$)}
          & (mag/(\arcsec)$^{2}$)	 \\
  (1) & (2) & (3) & (4) & (5) & (6) & (7) & (8) & (9) & (10) & (11) & (12) 
      & (13) & (14) & (15) & (16) & (17) & (18) & (19) \\
\hline
10	&	418.48	&	798.70	&	      12.14	&	1289.33	&	3107.21	&	10.44	&	1791.66	&	3262.49	&	9.81	&	9710.94	&	16982.44	&	7.39	&	12234.36	&	12432.62	&	4.43	&	1332.25	&	331.76	&	4.42	\\
20	&	68.84	&	41.95	&	     14.10	&	58.98	&	47.21	&	13.79	&	187.16	&	108.67	&	12.26	&	562.44	&	796.98	&	10.48	&	1884.21	&	2029.40	&	6.46	&	1470.84	&	757.50	&	4.32	\\
30	&	23.89	&	13.64	&	     15.25	&	18.32	&	11.90	&	15.06	&	61.18	&	35.09	&	13.47	&	146.61	&	95.55	&	11.94	&	514.92	&	664.68	&	7.87	&	1379.43	&	1100.25	&	4.39	\\
40	&	11.79	&	4.76	&	     16.01	&	8.63	&	4.19	&	15.87	&	28.21	&	14.93	&	14.31	&	60.26	&	36.95	&	12.91	&	138.81	&	145.15	&	9.29	&	1091.89	&	1001.56	&	4.64	\\
50	&	7.07	&	2.36	&	     16.57	&	4.95	&	1.95	&	16.48	&	14.94	&	8.39	&	15.00	&	26.54	&	17.62	&	13.80	&	66.10	&	53.09	&	10.10	&	834.47	&	860.51	&	4.93	\\
60	&	5.14	&	1.77	&	     16.91	&	3.50	&	1.26	&	16.85	&	9.23	&	4.41	&	15.53	&	14.95	&	8.82	&	14.42	&	41.26	&	33.90	&	10.61	&	600.75	&	651.90	&	5.29	\\
.	&	.	&		&	.	&	.	&	.	&	.	&	.	&	.	&	.	&	.	&	.	&	.	&	.	&	.	&	.	&	.	&	.	&	.	\\
.	&	.	&		&	.	&	.	&	.	&	.	&	.	&	.	&	.	&	.	&	.	&	.	&	.	&	.	&	.	&	.	&	.	&	.	\\
.	&	.	&		&	.	&	.	&	.	&	.	&	.	&	.	&	.	&	.	&	.	&	.	&	.	&	.	&	.	&	.	&	.	&	.	\\
\hline
\end{tabular}
Notes: Table~\ref{sbtable} is published in its entirety in the electronic 
  edition of the {\it MNRAS}. A portion is shown here for guidance regarding 
  its form and content.
\end{minipage}
\end{sidewaystable}

\section{Photometry\label{phot}}

We performed basic photometry and source characterization of the Circinus 
Galaxy by using the pipeline developed for the 2MASS Large Galaxy Atlas 
\citep{Jarrett03} and updated for the WISE High Resolution Galaxy Atlas 
(Jarrett et al. 2012). The pipeline provides an interactive system to identify 
foreground (contaminant) stars and assists in shape/extent characterization, 
surface brightness and integrated flux measurements. The user specifies the 
initial shape parameters, including the annulus used to estimate the local 
background and the elliptical aperture for source measurements. For each band, 
the local background is determined from the mode of the  pixel value distribution, a robust metric 
for the ``sky'', and thus is adopted as the local background value. 

The shape of the galaxy is determined using the 3\,$\sigma$ elliptical isophote, 
while the size or extent of the visible galaxy corresponds to the 1$\sigma$ 
isophote. The isophotal fluxes correspond to the integral of the elliptical 
shape defined by the axis ratio, position angle and isophotal radius. 
Additionally, in order to compare fluxes across bands to derive colors, we 
adopt the IRAC 8~$\mu$m aperture as the fiducial aperture for all IRAC 
measurements (whereas the MIPS 24~$\mu$m and 70~$\mu$m measurements are derived from their 
own isophotal apertures). Finally, the azimuthal elliptical-radial surface 
brightness is then characterized using a double S\'ersic function, where one 
component is fit to the inner region (i.e., the bulge) of the galaxy and the 
second component is fit to the outer region (i.e., the disk). The fit to the 
radial surface brightness is used to estimate the total flux by extrapolating 
the fit to larger radii, corresponding to three times the disk scale length 
beyond the isophotal radius. This threshold represents a practical balance 
between capturing the outer disk light and mitigating fit errors that 
accumulate with radius. 

The source characterization results are as follows.  The central position is determined
to be RA (J2000) = 213.29130\degr, Dec (J2000) = $-$65.33926\degr.
The 3\,$\sigma$ isophote shape is an ellipse with axis ratio $b/a$ = 0.485 
(ie. an inclination angle of $\sim$63\degr) 
at a position angle of $PA$ = 28.4\degr\ (208.4\degr).
The size of the galaxy based on the 1\,$\sigma$ (fiducial) 
isophote radius is $R_{\rm iso}$ = 603\farcs5 (20.1\arcmin\ in diameter).
Using these parameters,
we determine the integrated flux density, $F_{\lambda}$, in all MIR bands. 
Note that for the case of MIPS-70, an additional color correction is required
to account for the spectral differences between Circinus (whose MIR spectrum
has a power-law spectral index of -2) and the MIPS calibrators.  The color
correction, or scaling factor, is 1.09 (see the MIPS Instrument Handbook).

The results of the Circinus photometry are summarized in Table~\ref{phot}.
The photometric uncertainty incorporates the formal (Poisson) errors as well
as the estimated background uncertainty.  However, given that a significant foreground
stellar population was removed from the images (notably for 3.6 and 4.5~$\mu$m), 
and at longer wavelengths the Galactic ISM subtraction was a significant operation,
the quoted uncertainties are likely to be too generous and thus represent a lower limit.

\begin{table} 
\begin{minipage}{60mm}
\caption{{\it Spitzer} photometry results of the Circinus Galaxy. }
\label{phot}
\begin{tabular}{lcc}
\hline
\hline
   Flux density      & [Jy]             & [mag]            \\
\hline
$F_{\rm 3.6 \mu m}$  & $6.78  \pm 0.12$ & $4.04 \pm 0.02$ \\
$F_{\rm 4.5 \mu m}$  & $7.81  \pm 0.15$ & $3.41 \pm 0.02$ \\
$F_{\rm 5.8 \mu m}$  & $14.78 \pm 0.28$ & $2.23 \pm 0.02$ \\
$F_{\rm 8.0 \mu m}$  & $47.82 \pm 0.91$ & $0.32 \pm 0.02$ \\
$F_{\rm 24 \mu m}$   & $79.65 \pm 1.49$ & $-2.61 \pm 0.02$ \\
$F_{\rm 70 \mu m}$   & $280.0 \pm 4.9$ & $-6.38 \pm 0.02$ \\
\hline
\end{tabular}
Notes: The MIPS-70 integrated flux was measured using a circular aperture
of radius 10 arcmin.  The flux was then color corrected with a scaling factor of 1.09,
appropriate for a steeply rising MIR spectrum,
as specified in the MIPS Instrument Handbook, 
\end{minipage}
\end{table}

\section{Reddening and Spectral Energy Distribution\label{SED}}

Dust reddening affects the measured fluxes and is particularly severe for 
the Circinus Galaxy. A common method for estimating the reddening relies 
on the all-sky extinction map of \citet{DS98}, which is based on the {\it 
Infrared Astronomical Satellite} (IRAS), and {\it Diffuse Infrared Background 
Experiment} (DIRBE) measurements of IR dust emission. However, this map is 
inaccurate in high dust column density area and uncalibrated for $|b| <$ 
5\degr; it also tends to overestimate the reddening in regions of high 
extinction (see e.g., \citealp{AG99}). A higher spatial 
resolution all-sky dust emission map is needed to solve the problem of 
estimating the reddening in high extinction regions. A recent study 
by \citet{Kohyama10} has proposed a new method to construct a high spatial 
resolution extinction map but it is currently limited to the Cygnus region.  

Another way to derive the dust extinction towards Circinus is by comparing 
its near-IR (NIR) and MIR Spectral Energy Distribution (SED) to galaxy model 
templates with different values of $A_{\rm V}$. The model templates were
obtained from the SWIRE library \citep{Polletta07}, the majority of which 
were generated from the GRASIL code \citep{Silva98}. 
    
In order to obtain the Circinus SED with various $A_{\rm V}$ values, we adopt 
the wavelength-dependent extinction law of \citet{Cardelli89}:  
\begin{equation}
  A_{\lambda}/A_{\rm V} = a(x)+b(x)/R_{\rm V},   
\end{equation} 
where $a(x) = 0.574 x^{1.61}$, $b(x) = -0.527 x^{1.61}$, $x = 1/\lambda$ in 
$\mu$m$^{-1}$, and $R_{\rm V}$ = 3.1 for broadband filters. This relation 
was used to calculate the intrinsic magnitude, defined as $M_{\rm Intrinsic} 
= M_{\rm observed} - A_{\lambda}$, and then converted to flux with $F_\nu = 
F_{0} \times 10^{-0.4 M_{\rm Intrinsic}}$. The zero-magnitude flux, $F_0$, 
of Vega (see Section~4) is 1594.0, 1024.0, and 666.7~Jy for $J$, $H$, and 
$K_{s}$ bands, respectively. The shift from observed to rest wavelength is 
negligible at the redshift of Circinus ($z$ = 0.00145). We then fit the 
Circinus SED with various values of $A_{\rm V}$ to the best represented 
Seyfert\,2 model template. The template was normalized to the 2MASS $J$-band 
value. We derive a visual extinction of $A_{\rm V}$ = $2.1 \pm 0.4$ mag. 
Comparing our value to the $A_{\rm B}$ = 6.276 and $A_{\rm V}$ = 4.694 from 
\citet{SFD98}, we confirm the previous finding of over-estimation based on 
the current all-sky extinction map in high column density region.

In Figure~\ref{SED} we show the constructed Circinus SED using 2MASS $J$, $H$, 
and $K_{\rm s}$ values \citep{Jarrett03}, and our IRAC and MIPS 
measurements (black dots).  For comparison, we include the high-resolution
ISO-SWS spectrum (magenta) of the nuclear and circumnuclear regions of Circinus
\citep{Moorwood96}.  The spectrum contains a number of high excitation
ionic and molecular emission lines that arise from the strong radiation field that is 
sustained by the AGN.   Representing a composite of AGN and star formation features,
the spectral components that dominate the broad-band 
photometric measurements,
are the PAH bands at 6.2, 7.7, 8.6 and 11.3 $\mu$m, deep silicate absorption at 10 $\mu$m,
and the steeply rising continuum.    
Model templates are fit and normalized to the near-infrared,
including the a Seyfert 2 (red) and 
Seyfert 1.8 model templates (green), as well as
old stellar population, elliptical galaxy model template (blue)
to trace the host stellar light. The NIR 
region of the SED is dominated by the bulge population while the MIR region 
is dominated by the molecular and dust components of the Circinus ISM. 
We summarize our results in 
Table~\ref{cor_values}. 

\begin{figure*}
\includegraphics[width=17cm]{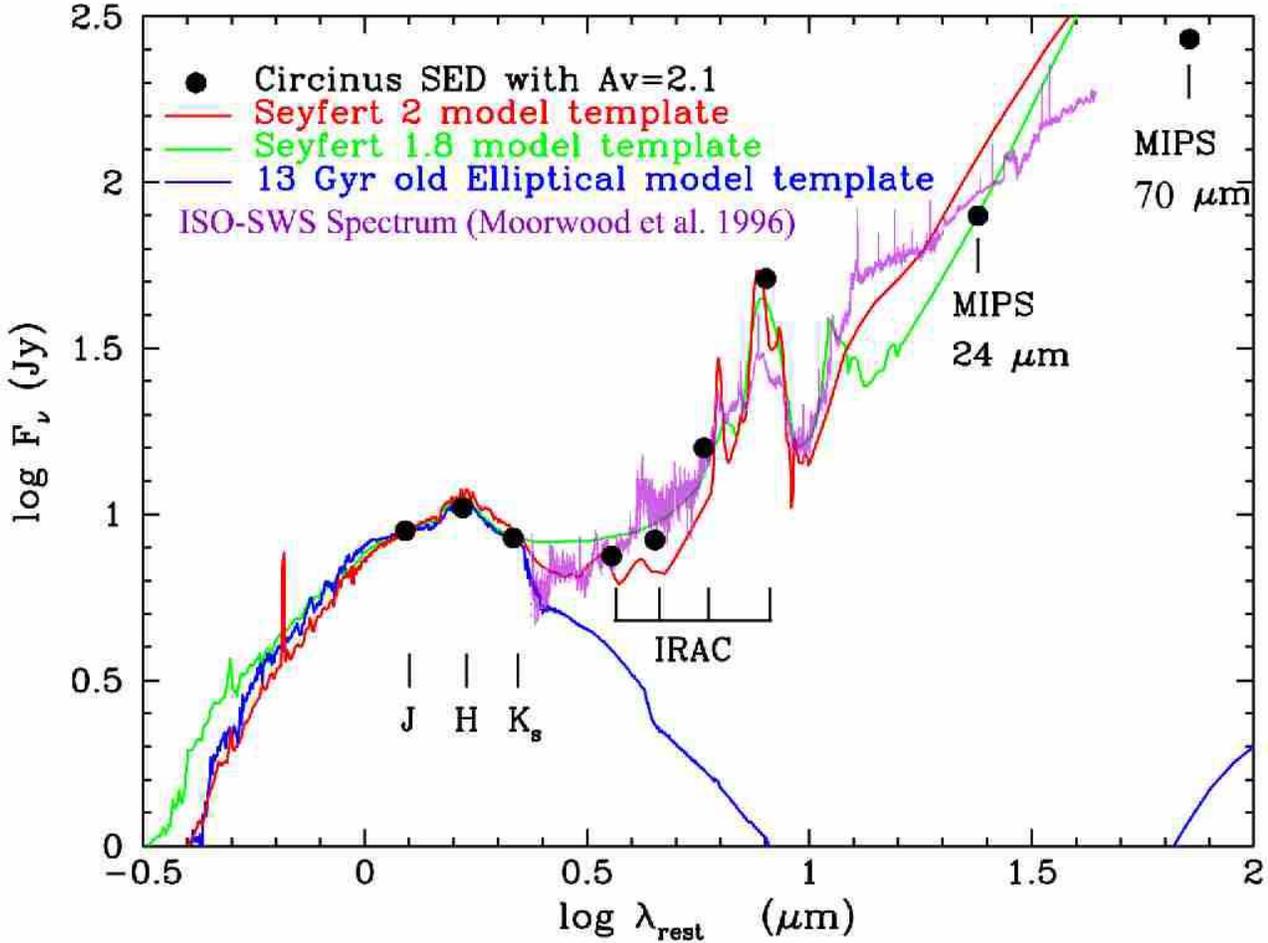}
\caption{The Circinus SED with 2MASS $J$, $H$, $K_{\rm s}$ band fluxes from 
   the Large Galaxy Atlas \citep{Jarrett03} and MIR measurements from the 
   IRAC and MIPS bands. Overlaid in magenta is the high resolution
   ISO-SWS spectrum of the central region of Circinus \citep{Moorwood96}.
      The galaxy model templates are retrieved from the 
   SWIRE library \citep{Polletta07} and normalized to the 2MASS $J$-band 
   value. We derive a visual extinction of $A_{\rm V}$ = 2.1~mag.}
\label{SED}
\end{figure*}

\begin{table}
\begin{minipage}{80mm}
\caption{Extinction-corrected total magnitudes ($M_{\rm Intrinsic}$), fluxes 
   ($F_{\nu}$) and spectral luminosities ($\nu L_{\nu}$) in different NIR 
   and MIR bands for the Circinus Galaxy. The extinction, $A_\lambda$, is 
   calculated using Eq~2 and $A_{\rm V}$ = 2.1~mag.} 
\label{cor_values}
\begin{tabular}{rrrrr}
\hline
\hline
$\lambda_{\rm rest}$ &
$A_{\lambda}$ &
$M_{\rm Intrinsic}$ &
$F_{\nu}$ &
$\nu L_{\nu}$ \\
 ($\mu$m) &  (mag)  & (mag)  &  (Jy)   & (10$^9$\Lsun)\\
  \hline
   1.233  &  0.605  &  5.63 &   8.92 & 11.90 \\
   1.660  &  0.376  &  4.98 &  10.46 & 10.39 \\
   2.156  &  0.246  &  4.74 &   8.49 & 6.49 \\
   3.595  &  0.107  &  3.94 &   7.49 & 3.43 \\
   4.493  &  0.076  &  3.33 &   8.37 & 3.07 \\
   5.792  &  0.076  &  2.15 &  15.85 & 4.51 \\
   7.988  &  0.076  &  0.24 &  51.25 & 10.58 \\
  23.965  &  0.000  & -2.61 &  79.35 & 5.46 \\
  71.42  &  0.000  & -6.38 &  280.0 &  6.51\\
\hline
\end{tabular}
\end{minipage}
\end{table}

\section{Masses\label{mass}}

The total mass of a galaxy consists of the visible stellar, dust and gas masses 
as well as dark matter. In the following subsections, we derive the visible 
mass components of the Circinus Galaxy. 

\subsection{Stellar Mass\label{smass}}

The IRAC 3.6 and 4.5~$\mu$m bands primarily trace the stellar light in the 
disk and bulge of a galaxy. We represent the stellar mass surface density with
\begin{equation}
  \Sigma_{*}(r) = C_{\lambda} \Upsilon_{\lambda} f_{\rm A,\lambda} 
                  \frac{F(r)}{F_{0}}\cos i,     
\end{equation}
where $C_{\lambda}$ is the conversion factor for the corresponding wavelength 
bands ($C_{\rm 3.6\mu m}$ = 0.196\Msun\,pc$^{-2}$ and $C_{\rm 4.5\mu m}$ = 
0.201\Msun\,pc$^{-2}$; \citealp{Oh08}), $\Upsilon_{\lambda}$ is the 
mass-to-light ratio in solar unit, 
$F(r)$ is the inclination-corrected flux density in Jy, 
$f_{\rm A,\lambda}$ is the aperture correction factor and 
$F_{0}$ is the zero-magnitude flux of Vega 
(the respective values for $f_{\rm A}$ and $F_0$ are given in Section 4). 
The $\lambda$ index indicates a wavelength dependency for these quantities, 
which are usually determined within the specific photometric band. 

The stellar mass-to-light ratio, $\Upsilon_{\lambda}$, depends on 
many other physical parameters, 
such as the initial mass function (IMF), the recent star formation 
history of the galaxy, etc. 
Determining $\Upsilon_{\lambda}$,
optically was difficult until a study by \citet{BdJ01}, who 
found a strong correlation between the stellar mass-to-light ratio and the 
integrated color of the stellar population. To determine $\Upsilon_{\lambda}$ 
of Circinus in the IRAC bands, we can utilize its color indices. 

We adopt Eq.~(8) of \citet{Westmeier11}, which describes the relation between 
a galaxy's stellar mass-to-light ratio and its 2MASS $J-K_{\rm s}$ color 
index. This equation was derived from the relations of \citet{BdJ01}. To 
incorporate the global color index of $J-K_{\rm s}$ to the IRAC bands, we 
used Eqs.~(6) and (7) of \citet{Oh08}, which were derived from the stellar 
population synthesis model. The final adopted equations to calculate the 
stellar mass-to-light ratio in the IRAC 3.6 and 4.5~$\mu$m bands are: 
\begin{equation}
  \Upsilon_{3.6 \mu m}=0.92\times10^{(1.434(J-K_{s})-1.380)}-0.05,
\end{equation}
\begin{equation}
  \Upsilon_{4.5 \mu m}=0.91\times10^{(1.434(J-K_{s})-1.380)}-0.08.
\end{equation}

Using the extinction corrected $J$ and $K_{\rm s}$-band magnitudes (see 
Table~\ref{cor_values}), we obtain $\Upsilon_{3.6 \mu m} = 0.679\pm0.070$ and 
$\Upsilon_{4.5 \mu m} = 0.641\pm0.060$. We then use Eq.~(3) to calculate the radial 
stellar mass surface density with the flux density of each band. The 
resulting total stellar mass is $9.5 \times 10^{10}$\Msun. A summary of 
the stellar surface mass densities, $\Sigma_{*}(r)$, is given in 
Table~\ref{sigma_masses}. 

\subsection{Gas Masses\label{gmass}}

We derive the radial \HI\ gas mass surface density, $\Sigma_{\rm HI}(r)$ 
(\Msun\,pc$^{-2}$), assuming the same orientation parameters as before 
($PA$ = 210\degr\ and $i$ = 65\degr), out to a radius of 2000\arcsec. We use
\begin{equation}
  \Sigma_{\rm HI}(r) = m_{\rm H} N_{\rm HI}(r) \cos i,
\end{equation}
where $m_{\rm H} = 1.674 \times 10^{-27}$~kg is the mass of a hydrogen atom
and $N_{\rm HI}$ is the \HI\ column density in cm$^{-2}$. We estimate the
\HI\ mass as \MHI\ = $\Sigma_{\rm HI}(r) \times$ area, which resulted in 
$6.6 \times 10^{9}$\Msun. The single-dish estimate of \MHI\ = $7 \times 
10^{9}$\Msun\ (\citealp{Freeman77,Henning00,K2004}) is, as expected, 
slightly higher. 

The distribution of cold molecular hydrogen gas (H$_2$) in galaxies is 
difficult to measure directly. A proxy, such as the CO line, is generally 
used as a tracer. To estimate the H$_2$ gas mass surface density, 
$\Sigma_{\rm H_2}$, of the inner disk of the Circinus Galaxy we use the 
CO(1--0) map provided by \citet{Curran08}. To determine $\Sigma_{\rm H_2}$, 
we first compute the column density via the CO to H$_2$ conversion factor, 
$X_{\rm CO}$. The latter is defined by the ratio of the H$_2$ column density, 
$N_{\rm H_2}$ in cm$^{-2}$, to the integrated CO intensity, $I_{\rm CO}$ 
in K\kms. It has been derived using various methods such as the virial mass 
method \citep{Solomon87} and $\gamma$-ray emission caused by the collision of 
cosmic-rays with hydrogen \citep{Bloemen86}. There appears to be no universal 
$X_{\rm CO}$ factor as it varies with the metallicity of the galaxy, UV 
radiation field and cosmic-ray density. Nevertheless, previous studies 
suggested that $X_{\rm CO}$ is accurate in the Milky Way disk (see 
\citealp{Solomon87}; \citealp{Hunter97}). To compare the total molecular gas 
mass with the study by \citet{Curran08}, we adopt the same $X_{\rm CO}$ 
factor of $2.3 \times 10^{20}$ cm$^{-2}$/(K\kms) \citep{Strong88} and 
a 36\% mass correction for Helium, $f_{\rm He}$ = 1.36. We use
\begin{equation}
  \Sigma_{\rm H_2}(r) = 2m_{\rm H}f_{\rm He}N_{\rm {H_2}}(r)\cos i~.
\end{equation}
The measurements of $I_{\rm CO}$ were carried out from 0\arcsec\ to 300\arcsec\ 
(the furthest detection in the CO(1--0) map). For the radial ring estimates we 
adopt $i$ = 65\degr\ and $PA$ = 214\degr. We determine $\Sigma_{\rm H_2}$ = 
1441\Msun\,pc$^{-2}$ and derive $M_{\rm H_2} = 2.1 \times 10^9$\Msun.

We calculate a total gas mass, $M_{\rm gas}$ = \MHI\ + M$_{\rm H_2}$, of about
$9 \times 10^9$\Msun. A summary of the measured gas mass surface density of 
Circinus is presented in Table~\ref{sigma_masses}. Figure~\ref{smass} shows 
the stellar and gas masses surface densities as a function of radius.

\begin{table}
\centering
\begin{minipage}{70mm}
\caption{Radial mass surface density of gas and stellar components for the 
   Circinus Galaxy. A position angle of $PA$ = 210\degr\ and an inclination 
   angle of $i$ = 65\degr\ were used to derive all quantities, with the 
   exception of CO for which we adopted $PA$ = 214\degr.}
\label{sigma_masses}
\begin{tabular}{rcrrrr}
\hline
\hline
\multicolumn{2}{c}{Radius} & $\Sigma_{*}^{3.6 \mu m}$	
                           & $\Sigma_{*}^{4.5 \mu m}$	
                           & $\Sigma_{\rm HI}$	
                           & $\Sigma_{\rm H_2}$	\\
(\arcsec) & (kpc)	   & \multicolumn{2}{c}{(\MMsun~pc$^{-2}$)}
                           & \multicolumn{2}{c}{(\MMsun~pc$^{-2}$)} \\
  \hline
 10 & 0.20 & 79133.2 & 366294.1&	5.34	&    239.8 	\\
 20 & 0.41 & 13017.7 & 16754.8	&	5.85	&    237.3    \\
 30 & 0.61 & 4517.9  & 5203.2	&	6.17	&    189.1    \\
 40 & 0.81 & 2229.7  & 2452.3	&	6.09	&    168.5    \\
 50 & 1.02 & 1336.9  & 1406.3	&	5.72	&    97.5     \\
 60 & 1.22 & 972.2   & 995.5	&	5.32	&    104.2    \\
 70 & 1.43 & 729.9   & 742.1	&	5.29	&    72.5     \\
 80 & 1.63 & 592.1   & 599.7	&	5.17	&    53.9     \\
 90 & 1.83 & 496.0   & 500.9	&	5.52	&    38.4     \\
100 & 2.04 & 428.7   & 429.8	&	5.69	&    43.0     \\
110 & 2.24 & 379.3   & 379.8	&	5.59	&    33.9     \\
120 & 2.44 & 339.8   & 342.9	&	5.32	&    29.7     \\
130 & 2.65 & 309.2   & 310.5	&	5.24	&    23.8     \\
140 & 2.85 & 274.6   & 277.0	&	4.99	&    14.4     \\
150 & 3.05 & 249.6   & 251.1	&	4.87	&    12.4     \\
160 & 3.26 & 218.4   & 219.9	&	4.97	&    13.1     \\
.   &  .   &  .      &     .    &         .     &        .      \\
.   &  .   &  .      &     .    &         .     &        .      \\
.   &  .   &  .      &     .    &         .     &        .      \\
\hline
\end{tabular}
Notes: Table~\ref{sigma_masses} is published in its entirety in the electronic 
   edition of {\it MNRAS}. A portion is shown here for guidance regarding its 
   form and content.
\end{minipage}
\end{table}

\begin{figure*}
\includegraphics[scale=0.5]{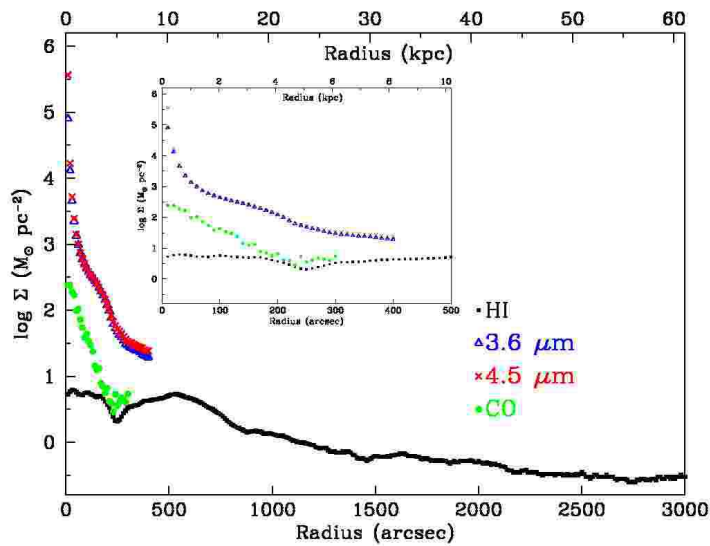}
\caption{The stellar and gas mass surface densities of the Circinus Galaxy. 
   The stellar component is derived from the IRAC 3.6 and 4.5~$\mu$m images 
   (within the 1$\sigma$ isophotal radius of 400\arcsec) and the gaseous 
   component from \HI\ and CO(1--0) maps (see Table~5). The inset highlights 
   our measurements at radii $<$ 500\arcsec\ (10~kpc).}
\label{smass} 
\end{figure*}

\section{Star Formation\label{SF}}

The interstellar medium (ISM) consists of dust, neutral gas and ionized gas, 
which are important ingredients to facilitate star formation (SF). As such, 
understanding the physical processes in the ISM and the star formation 
rate (SFR) are important to understand the physics of galaxy evolution and 
formation. 

Despite the complexity of the ISM--SFR relationship, a widely used 
parameterized power-law between the SFR surface density, $\Sigma_{\rm SFR}$, 
and the cold gas mass surface density, $\Sigma_{\rm gas}$, in the form
\begin{equation}
  \Sigma_{\rm SFR} \propto \Sigma_{\rm gas}^{N}, 
\end{equation} 
has been introduced by \citet{Schmidt59}. Later, observations and studies of 
a sample of spiral and starburst galaxies reveal a global correlation between 
the disk-averaged SFR and the gas mass surface density  (e.g., 
\citealp{Kennicutt98}; \citealp{KennicuttAR98}, and references therein), 
which is now known as the ``Kennicutt-Schmidt law'', with a power-law index 
of $N$ = 1.4. This measurement has been very useful as input to theoretical 
models of galaxy evolution (e.g., \citealp{Kay02}).

\subsection{The Global Star Formation Rate\label{GSFR}}

Star formation mostly occurs in the nuclear region and inner spiral arms of 
galaxies. Some star formation also occurs in their outer disks (e.g., M\,83; 
\citealp{Thilker05}; Koribalski et al. 2012, in prep.) and in interaction zones between 
galaxies (e.g., NGC 1512/1510 pair; \citealp{BA09}). Traditionally, the UV 
stellar continuum and optical nebular recombination lines have been used as 
SFR indicators but the former is severely affected by dust attenuation. On 
the other hand, MIR and FIR wavelengths are ideal to measure SFR due to UV 
photons from massive star formation absorbed by dust and re-radiated in these 
wavelengths. Numerous efforts to find SFR indicators related to MIR and FIR 
have been explored in the past decades (see e.g., \citealp{Calzetti94}; 
\citealp{Kennicutt98}; \citealp{Meurer99}; \citealp{Salim07}). For an 
extensive review on the subject of deriving SFR relationship from measurements 
at MIR and FIR wavelengths, and their caveats, see \citet{Calzetti07}. Here 
we use a range of SFR relations to estimate the obscured global SFR of the 
Circinus Galaxy. A summary of the calculations is given in Table~\ref{gsf}. 
We note that the PAH emission in the nuclear region is not related to the AGN 
as claimed in the NIR polarimetry study of the Circinus nucleus 
\citep{Alexander99}. We conclude that the obscured global SFR of the Circinus 
Galaxy is $\sim$3--8\Msun\,yr$^{-1}$. 
 
\begin{figure}
\includegraphics[scale=0.4]{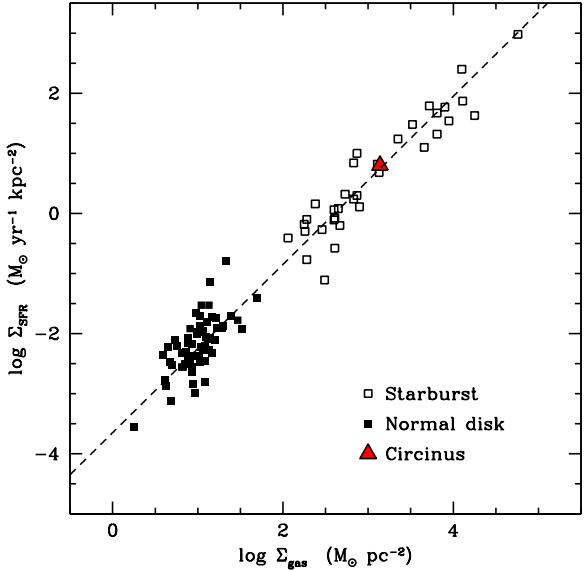}
\caption{Global Kennicutt-Schmidt law \citep{Kennicutt98} for normal disk 
   galaxies (filled squares) and starburst galaxies (open squares). The 
   Circinus Galaxy is marked with a red triangle.}
\label{sur_sfr} 
\end{figure}

In Figure~\ref{sur_sfr} we show the global Kennicutt-Schmidt relation for a 
sample of nearby star forming galaxies \citet{Kennicutt98}. The calculated 
logarithmic $\Sigma_{\rm SFR}$ of the Circinus Galaxy is indicated by a red 
triangle; its value ($\sim$0.8\Msun\,yr$^{-1}$\,kpc$^{-2}$) lies within the 
group of starburst galaxies. 

\begin{table}
\centering
\begin{minipage}{80mm}
\caption{Global star formation rates as derived from different calibration 
   methods for the Circinus Galaxy.}
\label{gsf}
\begin{tabular}{c|c|l|l}
\hline
\hline
      SFR           & Wavelength & Calibration Method & Ref. \\
(\MMsun\,yr$^{-1}$) &  Required  &                    & \\
\hline
4.7 & 70 $\mu$m & $L_{\rm TIR}$      & (7) \\
8.5 & 24$\mu$m & $L_{\rm H\alpha}$      & (1) \\
6.7 &  8$\mu$m & $L_{\rm H\alpha}$	      & (1) \\
8.2 & 24$\mu$m & $L_{\rm 1.4 GHz}$	      & (1) \\
7.6 &  8$\mu$m & $L_{\rm 1.4 GHz}$	      & (1) \\
2.8 & 24$\mu$m & $L_{\rm 24\mu m}$	              & (2) \\
4.3 & 24$\mu$m & $L_{\rm 24 \mu m}$	              & (3) \\
1.6 & 60 \& 100~$\mu$m & $L_{\rm FIR}$                & (4) \\
4.6 & 24$\mu$m & $L_{\rm 24 \mu m}$	              & (5) \\
0.7$^a$ & 1.4~GHz  & $L_{\rm 1.4 GHz}$                & (6) \\
3.5$^b$ & 1.4~GHz  & $L_{\rm 1.4 GHz}$                & (6) \\
\hline
\end{tabular}
Notes: $^a$ SFR for stellar masses $>$5\Msun\ and $^b$ for $>$0.1\Msun.
References: (1) \citet{Wu05}, (2) \citet{Calzetti07}, (3) \citet{Rieke09},
(4) \citet{Kennicutt98}, (5) \citet{AH06},  (6)  \citet{Condon02}
and (7) \citet{Zhu08}.
\end{minipage}
\end{table}

\subsection{The Local Star Formation Regions\label{LSFR}}

\begin{figure*}
\includegraphics[width=9.5cm]{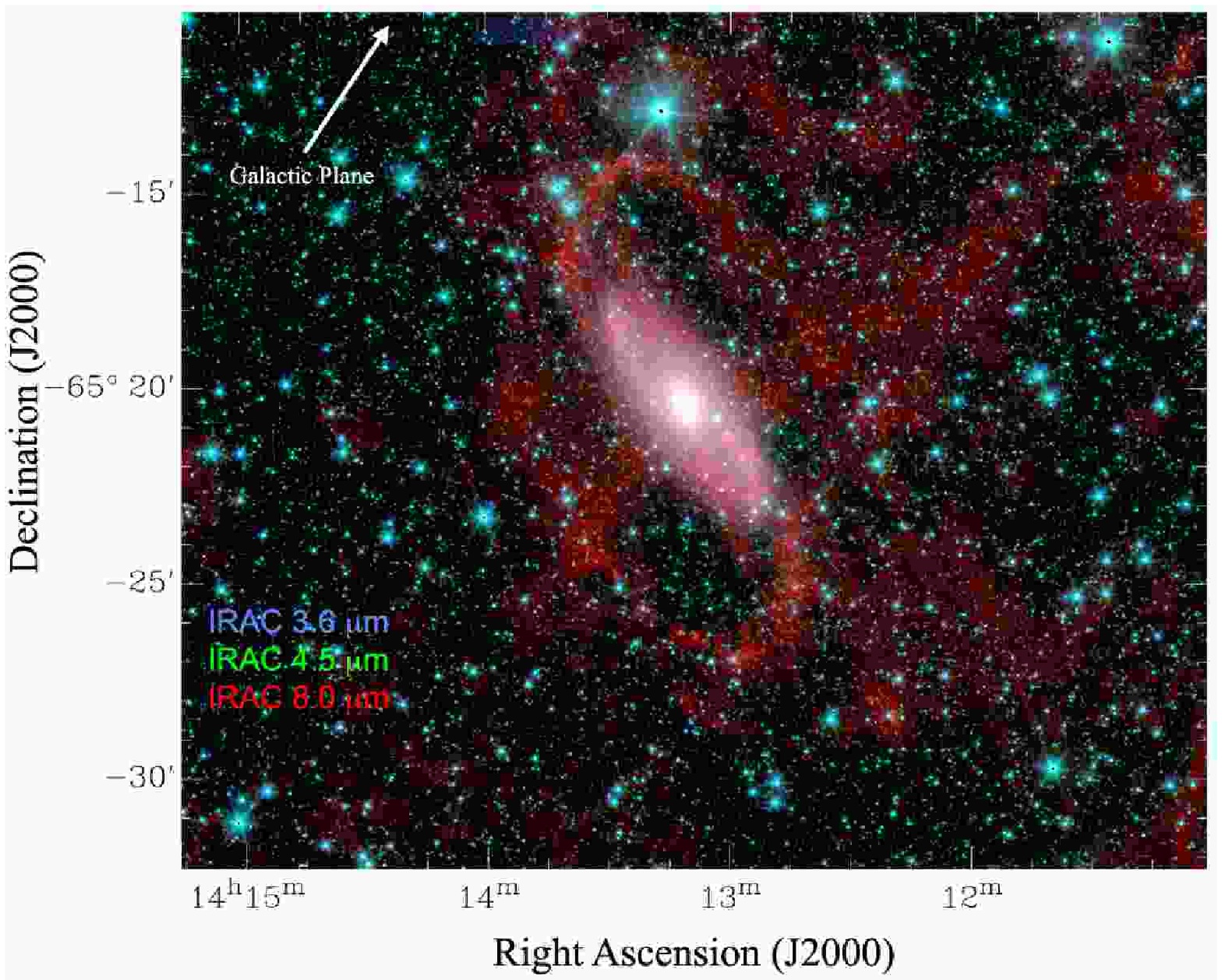} 
\includegraphics[width = 8cm]{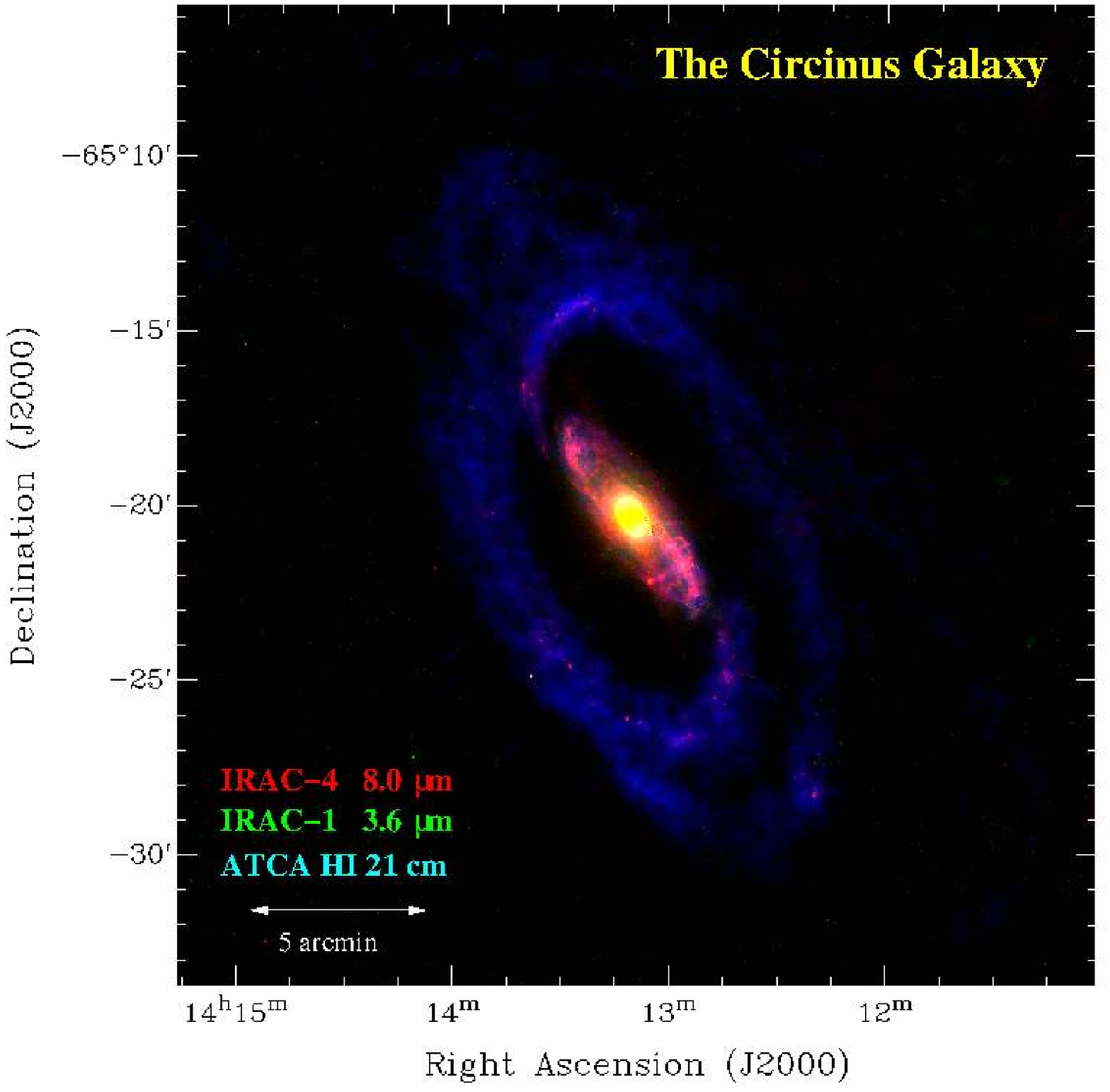}
\caption{Multi-wavelength color-composite image of the Circinus Galaxy. 
\textbf{Left:} {\it Spitzer} IRAC combination:  8~$\mu$m (red), 4.5~$\mu$m (green),
and 3.6$\mu$m (blue) images.  The large-scale Galactic emission and ``spur" (Fig. \ref{HIspur2}) have been removed
from the 8~$\mu$m image.
\textbf{Right:} A combination of the high-resolution ATCA \HI\ intensity map (blue) 
   and the {\it Spitzer} IRAC 8~$\mu$m (red) and 3.6~$\mu$m (green) images, 
   all three at a resolution of 15\arcsec.}
\label{composite_RGB} 
\end{figure*}

Gas provides an excellent reservoir for fueling star formation activities. 
This is particularly prominent in high gas density regions. In this section, 
we present the first attempt in identifying local star forming regions in the 
Circinus Galaxy by using \HI\ gas as a tracer. In Figure~\ref{composite_RGB}, 
we show a multi-wavelength color composite image of the Circinus Galaxy; two versions
are show:  the first is showcases the IRAC imaging, 3.6~$\mu$m (blue) + 4.5~$\mu$m (green)
+ 8.0~$\mu$m (red), and the second
consists of the high-resolution ATCA \HI\ map (blue), the IRAC 8~$\mu$m map
(red) and the IRAC 3.6~$\mu$m map (green). The physical size of the \HI\ gas 
distribution extends well beyond the area shown here. The multi-wavelength
composite images provide an powerful way to reveal the locations of star 
formation within the gaseous disk of Circinus.

\begin{figure*}
\includegraphics[scale=0.45]{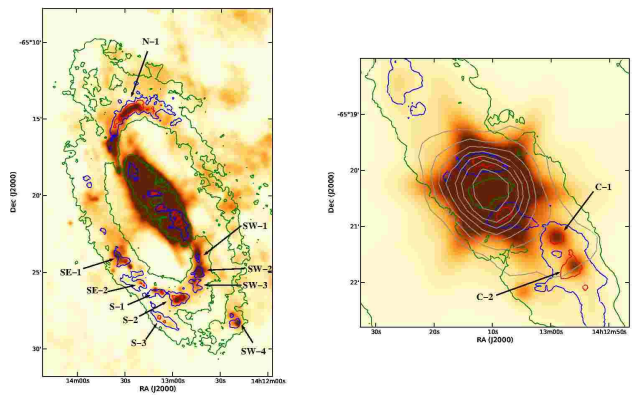}
\caption{Comparison of \HI, CO and mid-infrared emission in the inner disk of 
   the Circinus Galaxy. The high-resolution ATCA \HI\ intensity map (contours) 
   is superimposed onto the calibrated and foreground-subtracted {\it Spitzer} 
   IRAC 8.0~$\mu$m (left) and MIPS 24~$\mu$m (right) maps. Both {\it Spitzer} 
   maps have been convolved to 15\arcsec\ resolution to match the ATCA \HI\ 
   intensity map. The contour levels correspond to \HI\ column densities, 
   $N_{\rm HI}$, of 1.23 (green), 2.45 (blue) and 3.19 (red) $\times10^{21}$ 
   cm$^{-2}$. CO(1--0) contours (grey) have been added to the {\it Spitzer} 
   24~$\mu$m map (right) which shows only the central region. The labels 
   indicate local star forming regions as identified in Section~8.2.} 
\label{ch4_24_HI_comp}
\end{figure*}

In Figure~\ref{ch4_24_HI_comp}, we overlay the high-resolution ATCA \HI\ map 
in contours onto the IRAC 8~$\mu$m (left panel) and MIPS 24~$\mu$m (right 
panel) images. The contours start at \HI\ gas column densities of $1.23 \times 
10^{21}$ cm$^{-2}$. We identify 10 individual star forming regions in the IRAC 
8~$\mu$m map that are clearly associated with high density \HI\ gas. They are 
labeled according to their position with respect to the center of the Circinus 
Galaxy. The right panel shows the same \HI\ contours overlaid onto the MIPS 
24~$\mu$m image to highlight two resolved SF regions, C-1 and C-2, near the 
central region. For comparison we added CO(1--0) contours in grey; the map is
rather small and has an angular resolution of 50\arcsec\ \citep{Curran08}. 
The higher resolution (25\arcsec) CO(2--1) map (not shown here) reveals 
molecular gas concentrations at the same location as the high-density \HI\
clumps and clearly associated with the SF locations.

\begin{figure*}
\begin{tabular}{cc}
\includegraphics[scale=0.4]{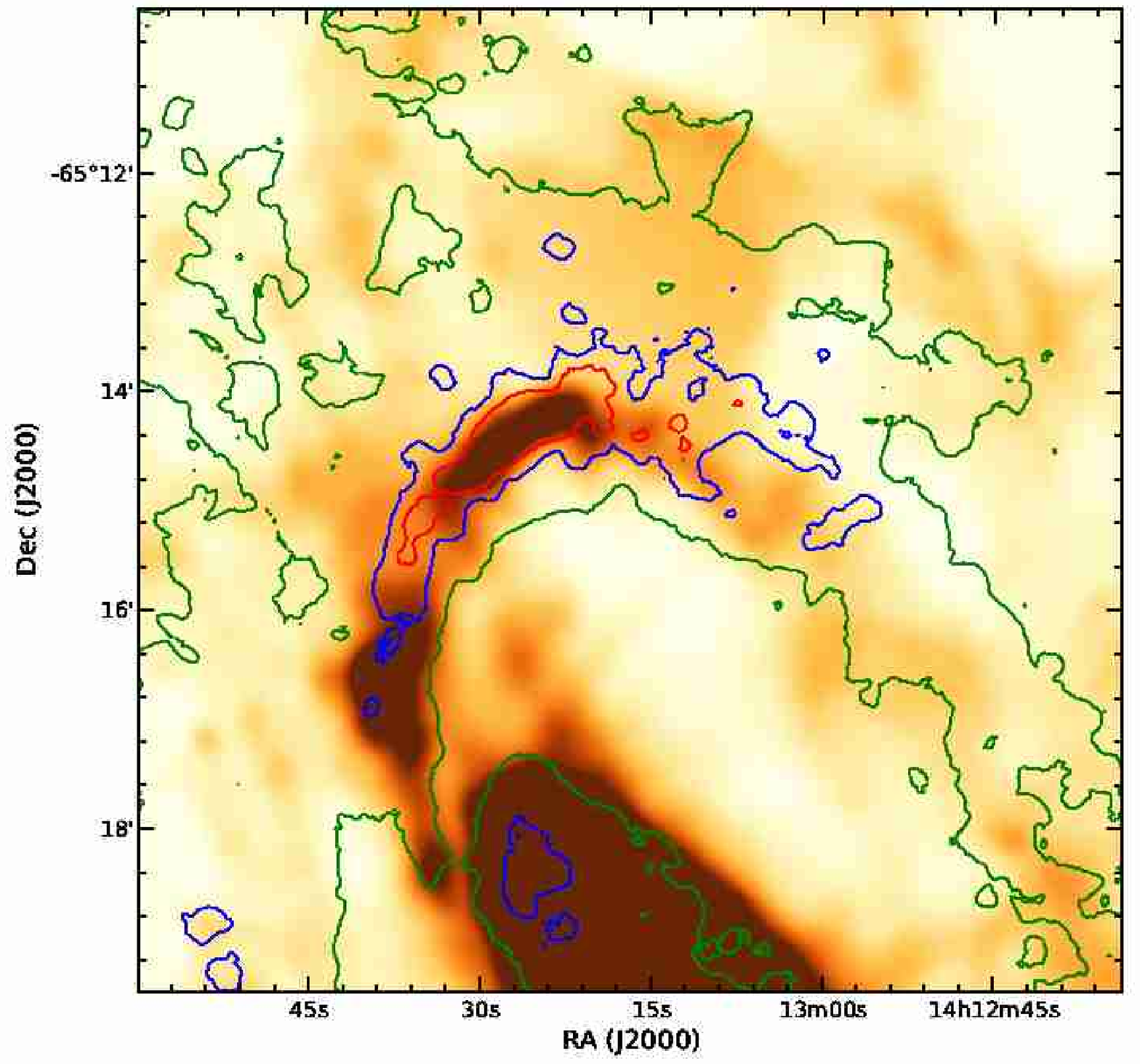} &
\includegraphics[scale=0.4]{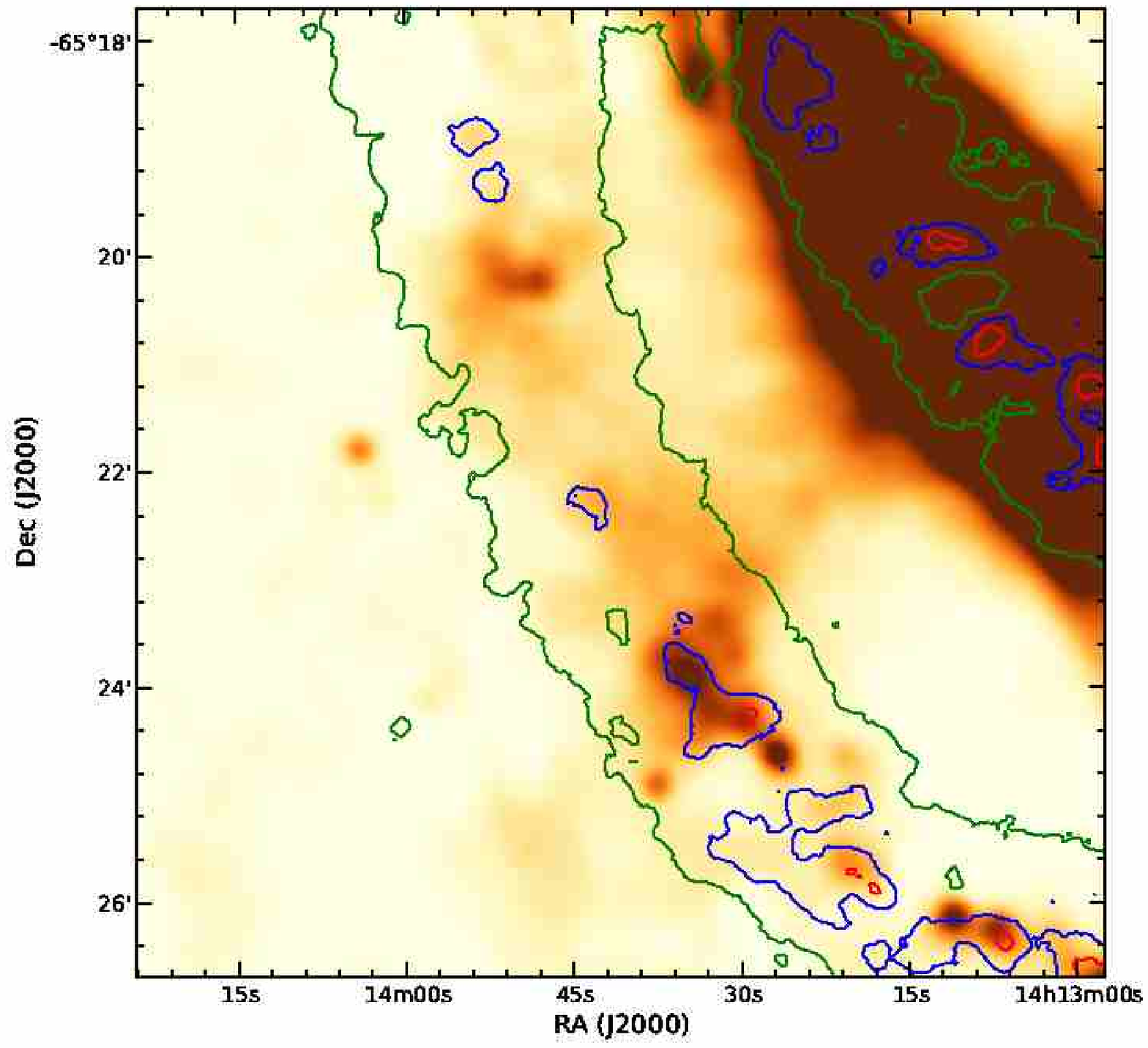} \\
\end{tabular}
\caption{Comparison of ATCA \HI\ and {\it Spitzer} 8~$\mu$m emission in the 
  inner spiral arms of the Circinus Galaxy. Here we show an enlarged view of 
  two different sections from Figure~\ref{ch4_24_HI_comp} (left panel). 
  While bright star forming regions typically coincide with high density \HI\ 
  emission, faint star forming regions appear offset to \HI\ enhancements.}
\label{mismatches} 
\end{figure*}

While there is generally a tight correlation between regions of high density 
gas and star formation activity (proportional to the MIR luminosity), we 
occasionally find exceptions (e.g., see Figure~\ref{mismatches}). Such 
mismatches have also been observed in other spiral galaxies
(e.g., M\,83 bar transition region as explored in Jarrett et al. 2012). A ready 
explanation is related to the Lin--Shu density wave theory \citep{LS64,LS66}, 
which was originally proposed to explain the spiral arm structure of galaxies. 
The idea suggests that the gas enters the density wave and is compressed 
due to enhanced gravity, which results in an increased local density. When 
the dense gas clouds reach the Jeans instability, they collapse to form new 
stars. As such, the local star formation rate in the region is also increased. 
Subsequently, the stars move out of the gas clouds, leaving the gas behind. 

To evaluate if the mismatches follow the density wave theory prediction, we 
need to know the rotation direction of the Circinus Galaxy (clockwise or
anti-clockwise). Based on the {\it Spitzer} composite image 
(see Figure~\ref{composite_RGB} and 
the \HI\ velocity field \citealp{Jones99}) we conclude that Circinus rotates 
anti-clockwise, i.e., is co-rotating with the density wave. This implies the 
stars should form and lead ahead of the gas clumps in an anti-clockwise 
direction. This also applies for all SF regions. 
Examining all the mismatches, we conclude 
that most of them are consistent with the prediction.

Many studies are either supporting or confronting the density wave theory. 
We highlight a recent study by \citet{Foyle11}, in which multi-wavelength 
data of 12 nearby spiral galaxies have been used to calculate the angular 
cross-correlations in different star formation sequence. They conclude that 
the density wave theory is an over-simplification in explaining the spiral 
arm structure as seen in large disk galaxies. 

\section{Comparison between 8 and 24 micron emission\label{ratio}}

A study by \citet{Bendo08} has shown that the relation between the stellar 
continuum-subtracted PAH 8~$\mu$m emission and 24~$\mu$m hot dust emission 
exhibits a large scatter within the central 2~kpc of galaxies. Strong spatial 
variations of the (PAH 8~$\mu$m) / 24~$\mu$m surface brightness ratio are 
also observed, with high values in the diffuse ISM and low values in bright 
star-forming regions. In order to study the (PAH 8~$\mu$m) / 24~$\mu$m 
surface brightness ratio in Circinus, we prepared the data in a similar 
manner as described in \citet{Bendo08}. In the following we briefly describe 
the procedure and our results. 

\begin{figure*}
\begin{tabular}{cc}
\includegraphics[scale=0.45]{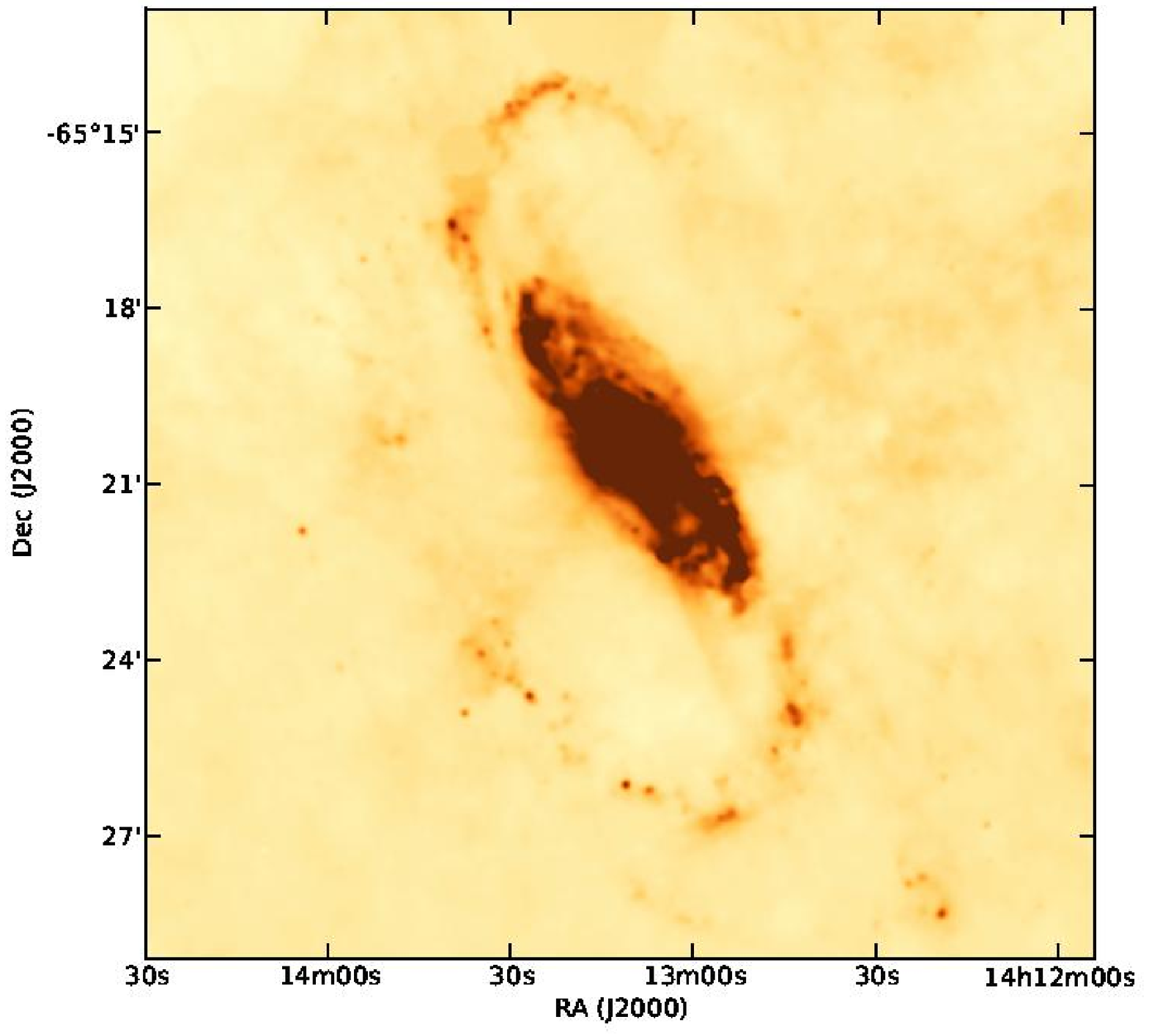} &
\includegraphics[scale=0.44]{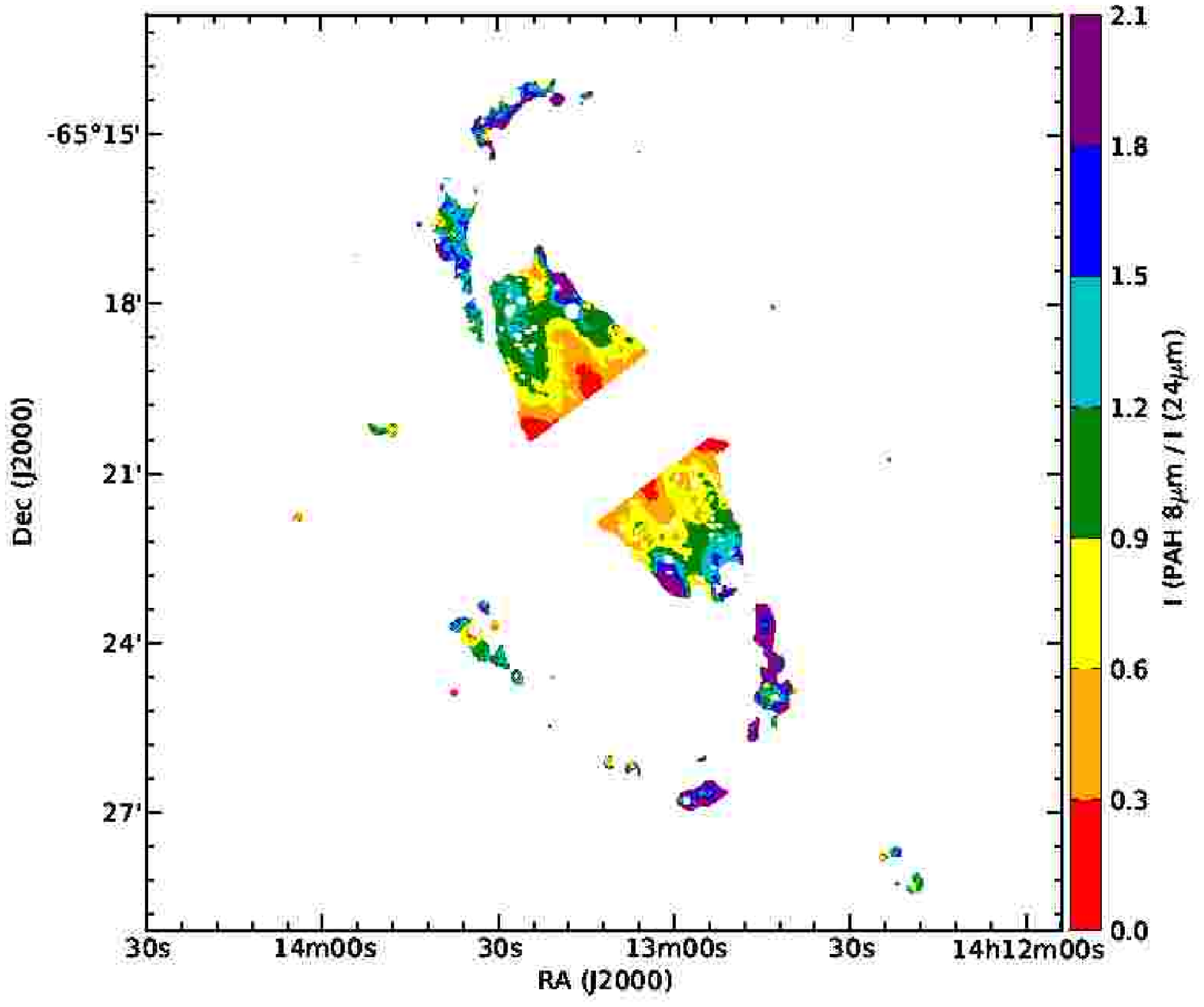}\\
\end{tabular}
\caption{{\it Spitzer} 8~$\mu$m image (left) and stellar continuum-subtracted 
   (PAH 8~$\mu$m) / 24~$\mu$m surface brightness ratio (right) of the Circinus 
   Galaxy, both at an angular resolution of 6\arcsec. The latter has been 
   obtained following \citep{Bendo08}; for comparison we use the same range
   for the ratio match. There is a clear color gradient from the center of
   Circinus to the edge of the inner disk and start of the spiral arms. 
   Artifacts affected by the bright core and foreground stars have been removed. 
   The red color 
   in the ratio map corresponds to relatively weak emission from PAH 8~$\mu$m 
   while blue indicates strong PAH emission. }  
\label{8to24_ratio} 
\end{figure*}

We first convolve the IRAC 3.6 and 8~$\mu$m images to match the PSF of the 
MIPS 24~$\mu$m image, which has a FWHM of 6\arcsec. The convolution is 
necessary to allow a direct comparison of surface brightness within the 
same aperture in different wavebands. The IRAC 3.6 and 8~$\mu$m images were 
then corrected for the diffusion of light through the IRAC detector. The 
effective aperture correction factors for these wavebands are listed in Section~4. 
Finally, we 
subtracted the stellar continuum from the IRAC 8~$\mu$m and MIPS 24~$\mu$m 
images using the equations derived by \citet{Helou04}

\begin{equation}
  I_{\rm PAH~8\mu m}  = I_{\rm 8\mu m}  - 0.232~I_{\rm 3.6\mu m}
\end{equation}
\begin{equation}
  I_{\rm 24-3.6\mu m} = I_{\rm 24\mu m} - 0.032~I_{\rm 3.6\mu m}.
\end{equation}
This results in a correction of about 30\% and 2\% to the 8 and 24~$\mu$m
surface brightness densities, respectively.
Subsequently, we took the ratio of $I_{\rm PAH~8\mu m}$ and $I_{\rm 24-3.6\mu m}$. 
Foreground stars, artifacts related to the bright galaxy core and 
low signal-to-noise areas ($<$3~$\sigma$) were excluded from the analysis.
The result is presented in Figure~\ref{8to24_ratio}. 

We find that the surface brightness ratio varies significantly across the 
disk of the Circinus Galaxy, with stronger point-like 24~$\mu$m emission 
among the identified local star forming regions and with PAH 8~$\mu$m emission 
dominating the diffuse ISM. The result is consistent with Bendo's finding on 
other galaxies. In contrast, the PAH 8~$\mu$m emission is dominant in the 
outer disk and along part of the spiral arms. The ratio variation of Circinus 
is similar to that of the nearby spiral galaxy NGC~3031 (see Figure~1 of 
\citealp{Bendo08}).   

\section{Summary and Conclusions\label{concl}}

We present a detailed study of the gas distribution and star formation 
activity in the inner and outer disk of the Circinus Galaxy. 
For this work, we utilized the infrared images 
obtained by the {\it Spitzer} IRAC and MIPS instruments, ATCA 
and Parkes \HI\ radio data, as well as SEST CO maps. 
Due to the location of Circinus behind the 
Galactic Plane, removal of Milky Way stars and Galactic emission from the 
MIR images was necessary prior to further analysis. The detection and 
characterization tools developed by WDSC and IPAC successfully removed the 
MW stars in all {\it Spitzer} images.  A correlation method and structure 
matching based on \HI\ gas in different velocity ranges were used to remove 
both large- and small-scale foreground structure from the IRAC 8~$\mu$m 
and MIPS 24~$\mu$m images. We demonstrate that 
the single-dish and interferometer combined Galactic \HI\ map can be
used to remove the large-scale foreground Galactic ISM.  Accordingly,
a scaled high resolution Galactic \HI\ map within a certain velocity range 
was used to remove the small-scale structure, notably the spur feature 
that, in projection, pierces the disk of the Circinus Galaxy. We confirm that the spur feature 
is related to the Galactic gas and with an estimated distance of 5~kpc. 
A detailed description of removing the Galactic ISM is given in Section~3.2. 

The physical properties via photometric analysis of the 
{\it Spitzer} images were determined for Circinus. 
Surface brightness profiles 
of Circinus in different 2MASS, IRAC and MIPS bands are shown in Figure 7. 
Using the spectral energy distribution, we derive a foreground visual extinction 
($A_{V}$) of 2.1 mag toward the Circinus Galaxy, which is notably lower than the value 
implied from the extinction map of \citet{SFD98}. 
Stellar mass-to-light ratios in 3.6~$\mu$m 
and 4.5~$\mu$m bands were determined and utilized to infer the total 
stellar mass of $9.5\times10^{10}$~\Msun. 
The sum of neutral hydrogen and molecular hydrogen 
masses gives a total gas mass of $9\times10^{9}$~\Msun for Circinus.

Using various wavelength calibrations, we derived the obscured star formation rates 
of the Circinus Galaxy between 3 and 8~\Msun~yr$^{-1}$. The star formation surface density 
follows the Kennicutt-Schmidt law and the value is similar to other 
starburst galaxies. Multiwavelength color composite images of Circinus 
as shown in Figure~11 reveal the prominent spiral arm, inner and outer disk. The 
\HI\ gas traces the PAH 8~$\mu$m feature, which strongly correlates with the 
local star forming regions. 
Cross-correlating the IRAC 8~$\mu$ and MIPS 24~$\mu$m 
images with the high-resolution ATCA \HI\ map, we identified 12 
distinct/resolved individual star forming regions. One of the star formation 
regions (SW-4) is located in the outskirt of Circinus. We also identified 
several offsets between the gas clumps and stars in Circinus. Analysis of 
these offsets supports the density wave scenario. However, a different 
scenario is likely needed to explain several anti-correlated exceptions. 

Finally, we compared the PAH 8~$\mu$m and 24~$\mu$m dust emission and found 
a significant variation across the disk of Circinus. Enhanced 24~$\mu$m 
emission relative to PAH 8~$\mu$m emission was also found in identified local star 
forming regions.  

\section*{Acknowledgments}

B.-Q. For is the recipient of a John Stocker Postdoctoral Fellowship from 
Australia's Science and Industry Endowment Fund (SIEF). She thanks the 
generous travel support to Australia from the University of Texas astronomy 
department Cox Excellence fund and the financial support by CSIRO Astronomy 
and Space Science during summer 2010. We also thank Michelle Cluver and 
Tobias Westmeier for helpful discussions.  
This work is based [in part] on observations made with the {\it Spitzer} and 
research using the  NASA/IPAC Extragalactic Database (NED) and
IPAC Infrared Science Archive,
all are operated by JPL, Caltech under a contract with the 
National Aeronautics and Space Administration.
Support for this work was provided by NASA through an award issued by JPL/Caltech.


\bibliographystyle{mn2e}
\bibliography{ref}

\end{document}